\documentclass[preprint]{aastex631}

\usepackage{savesym}
\savesymbol{tablenum}
\usepackage{siunitx}
\restoresymbol{SIX}{tablenum}
\usepackage{physics}
\usepackage{tikz}
\usetikzlibrary{shapes}
\usepackage{xcolor}
\usepackage[caption = false]{subfig}

\newcommand{\vFR}{\vb{v}_\mathrm{FR}}
\newcommand{\vHT}{\vb{v}_\mathrm{HT}}
\newcommand{\avg}{\expval}

\newcommand{\alf}{Alfv$\acute{\text{e}}$n} %plasma people use {\alf} a lot
\newcommand{\wal}{Wal$\acute{\text{e}}$n}

\submitjournal{ApJS}

\begin{document}

\title{A Closer Look at Small-Scale Magnetic Flux Ropes in the Solar Wind at 1 AU: Results from Improved Automated Detection}

\correspondingauthor{Hameedullah Farooki}
\email{haf5@njit.edu}

\author[0000-0001-7952-8032]{Hameedullah Farooki}
\affiliation{Institute for Space Weather Sciences,
New Jersey Institute of Technology,
University Heights, Newark, NJ, USA}
\affiliation{Center for Solar-Terrestrial Research,
New Jersey Institute of Technology,
University Heights, Newark, NJ, USA}

\author[0000-0002-8032-7833]{Sung Jun Noh}
\affiliation{Los Alamos National Laboratory, Los Alamos, NM}

\author[0000-0002-5865-7924]{Jeongwoo Lee}
\affiliation{Institute for Space Weather Sciences,
New Jersey Institute of Technology,
University Heights, Newark, NJ, USA}
\affiliation{Center for Solar-Terrestrial Research,
New Jersey Institute of Technology,
University Heights, Newark, NJ, USA}

\author[0000-0002-5233-565X]{Haimin Wang}
\affiliation{Institute for Space Weather Sciences,
New Jersey Institute of Technology,
University Heights, Newark, NJ, USA}
\affiliation{Center for Solar-Terrestrial Research,
New Jersey Institute of Technology,
University Heights, Newark, NJ, USA}
\affiliation{Big Bear Solar Observatory, New Jersey Institute of Technology, Big Bear City, CA, USA}

\author[0000-0002-6350-405X]{Hyomin Kim}
\affiliation{Institute for Space Weather Sciences,
New Jersey Institute of Technology,
University Heights, Newark, NJ, USA}
\affiliation{Center for Solar-Terrestrial Research,
New Jersey Institute of Technology,
University Heights, Newark, NJ, USA}

\author[0000-0003-0792-2270]{Yasser Abduallah}
\affiliation{Institute for Space Weather Sciences,
New Jersey Institute of Technology,
University Heights, Newark, NJ, USA}
\affiliation{Department of Computer Science,
New Jersey Institute of Technology, University Heights, Newark, NJ, USA}

\author[0000-0002-2486-1097]{Jason T. L. Wang}
\affiliation{Institute for Space Weather Sciences,
New Jersey Institute of Technology,
University Heights, Newark, NJ, USA}
\affiliation{Department of Computer Science,
New Jersey Institute of Technology, University Heights, Newark, NJ, USA}

\author[0000-0002-0065-7622]{Yu Chen}
\affiliation{Center for Space Plasma and Aeronomic Research (CSPAR), The University of Alabama in Huntsville, Huntsville, AL, USA}

\author[0000-0001-8184-2151]{Sergio Servidio}
\affiliation{Dipartimento di Fisica, Universita della Calabria, I-87036 Cosenza, Italy}

\author[0000-0003-4168-590X]{Francesco Pecora}
\affiliation{Department of Physics and Astronomy, University of Delaware, Newark, DE, USA}
%% Note that the \and command from previous versions of AASTeX is now
%% depreciated in this version as it is no longer necessary. AASTeX 
%% automatically takes care of all commas and "and"s between authors names.

%% AASTeX 6.31 has the new \collaboration and \nocollaboration commands to
%% provide the collaboration status of a group of authors. These commands 
%% can be used either before or after the list of corresponding authors. The
%% argument for \collaboration is the collaboration identifier. Authors are
%% encouraged to surround collaboration identifiers with ()s. The 
%% \nocollaboration command takes no argument and exists to indicate that
%% the nearby authors are not part of surrounding collaborations.

%% Mark off the abstract in the ``abstract'' environment. 
\begin{abstract}

Small-scale interplanetary magnetic flux ropes (SMFRs) are similar to ICMEs in magnetic structure, but are smaller and do not exhibit ICME plasma signatures.
We present a computationally efficient and GPU-powered version of the single-spacecraft automated 
SMFR detection algorithm
based on the Grad-Shafranov (GS) technique. Our algorithm is capable of processing higher 
resolution data, eliminates selection bias caused by a fixed $\avg{B}$ threshold,
has improved detection criteria demonstrated to have better results on an MHD simulation, and recovers full 
2.5D cross sections using GS reconstruction.
We used it to detect 512,152 SMFRs from 27 years (1996 to 2022)
of 3-second cadence \emph{Wind} measurements. Our novel findings are: 
(1) the radial density of SMFRs at 1 au (${\sim}1$ per $\si{10^6\kilo\meter}$) and filling factor (${\sim}$35\%) are independent of solar activity,
distance to the heliospheric current sheet (HCS),
and solar wind plasma type,
although the minority of SMFRs with diameters greater than ${\sim}$0.01 au have a strong solar activity dependence;
(2) SMFR diameters follow a log-normal distribution that peaks below the resolved range ($\gtrsim 10^4$ km), although
the filling factor is dominated by SMFRs between $10^5$ to $10^6$ km;
(3) most SMFRs at 1 au have strong field-aligned flows like those from PSP measurements;
(4) in terms of diameter $d$, SMFR poloidal flux $\propto d^{1.2}$, axial flux 
$\propto d^{2.0}$,
average twist number $\propto d^{-0.8}$,
current density $\propto d^{-0.8}$,
and helicity $\propto d^{3.2}$.
Implications for the origin of SMFRs and switchbacks are briefly discussed.
The new algorithm and SMFR dataset are made freely available.
\end{abstract}

%% Keywords should appear after the \end{abstract} command. 
%% The AAS Journals now uses Unified Astronomy Thesaurus concepts:
%% https://astrothesaurus.org
%% You will be asked to selected these concepts during the submission process
%% but this old "keyword" functionality is maintained in case authors want
%% to include these concepts in their preprints.

%% From the front matter, we move on to the body of the paper.
%% Sections are demarcated by \section and \subsection, respectively.
%% Observe the use of the LaTeX \label
%% command after the \subsection to give a symbolic KEY to the
%% subsection for cross-referencing in a \ref command.
%% You can use LaTeX's \ref and \label commands to keep track of
%% cross-references to sections, equations, tables, and figures.
%% That way, if you change the order of any elements, LaTeX will
%% automatically renumber them.
%%
%% We recommend that authors also use the natbib \citep
%% and \citet commands to identify citations.  The citations are
%% tied to the reference list via symbolic KEYs. The KEY corresponds
%% to the KEY in the \bibitem in the reference list below. 

\section{Introduction} \label{sec:intro}

It is well-known that the solar wind's magnetic field is only well described by the Parker spiral model on average; at a given point in time, fluctuations deflect the measured magnetic field away from the Parker spiral prediction.
Often, the magnetic field fluctuations are correlated with velocity fluctuations, so most early studies viewed them as non-interacting transverse {\alf} waves (e.g. \citet{belcher_large-amplitude_1971}).
However, observations showed that solar wind fluctuations, even when {\alf}ic, usually exhibit signatures not consistent with pure {\alf} waves, suggesting the presence of nonpropagating structures advected with the rest frame of the solar wind
\citep{burlaga_micro-scale_1968, burlaga_macro-_1968,burlaga_magnetic_1970,burlaga_microscale_1976,denskat_multispacecraft_1977,burlaga_pressure-balanced_1990,matthaeus_evidence_1990}.
Various theories of solar wind fluctuations consisting of a combination of advected structures and waves were formulated
(such as \citet{tu_model_1993}).
A popular version of this idea was put forth by \citet{borovsky_flux_2008}, wherein the solar wind is considered
a sea of magnetic flux tubes that are non-evolving fossil structures originating at the Sun's surface, or strands of the magnetic carpet. In this picture, flux tubes walls correspond to the observed discontinuities in the solar wind and turbulence is restricted to within the flux tubes.
However, an alternative possibility is local generation via magnetohydrodynamic (MHD) turbulence:
the cascade of helicity to large scales
can form twisted flux ropes in less than the time it takes the solar wind to propagate to 1 au \citep{matthaeus_spectral_2007,greco_intermittent_2008,servidio_depression_2008,greco_waiting-time_2009,wan_generation_2009,zank_theory_2017}.

Small-scale magnetic flux ropes (SMFRs) were first reported by
\citet{moldwin_ulysses_1995,moldwin_small-scale_2000},
described as transients with magnetic field signatures consistent with flux ropes observed interplanetary coronal mass ejections (ICMEs),
but without ICME plasma signatures such as reduced temperature.
Various studies have been performed on SMFRs since then, but they
were all based on very small lists of events, a few hundred at most.
The origin of these structures was debated, with two key possibilities being local reconnection across the heliospheric current sheet (HCS) \citep{moldwin_small-scale_2000,cartwright_heliospheric_2010} or small CMEs from the Sun \citep{feng_interplanetary_2008}.
A significant advancement was made when
\citet{zheng_automated_2017} introduced an automated detection 
algorithm based on the Grad-Shafranov (GS) technique (for a brief overview of the GS technique, which plays an important role in this work, see Appendix~\ref{sec:theory}). The detection algorithm was applied to produce a catalog of 74,241 SMFRs over 21 years of \emph{Wind} measurements
(\citet{hu_automated_2018}; referred to as the ``original catalog'', generated with the ``original algorithm'', hereafter).

Considering the large number of SMFRs, are they really transient structures, or are they an essential component of the solar wind?
\citet{hu_automated_2018} pointed out that
their results were consistent with considering the solar wind as a sea of flux tubes due to the abundance of SMFRs (${\sim}25\%$
of the time is contained in an SMFR in their catalog).
Recently, in a systematic study SMFR properties using machine learning, we found 
that there is essentially no difference between SMFRs and 
``background'' solar wind other than
differences imposed by the fixed $\avg{B} > \SI{5}{\nano\tesla}$ threshold in the original algorithm
(Farooki et al. 2023; manuscript submitted to ApJ).
Similarly, \citet{zhai_properties_2023}
found that the properties of most SMFRs are the same as the properties of the background solar wind.
Although these observations suggest that SMFRs are not transients, this is complicated by the observation of a strong solar cycle and HCS proximity dependence for SMFRs in the original catalog, seemingly consistent with the hypotheses of generation via reconnection across the HCS, small CMEs, or even solar eruptions that travel with the HCS as their conduit \citep{higginson_structured_2018}.

Application of the GS-based automated detection algorithm
to data from the Parker Solar Probe (PSP)
\citep{chen_small-scale_2020,chen_small-scale_2021,chen_small-scale_2022}
has shown that static SMFRs with low {\alf}icity (correlation between velocity and magnetic field fluctuations)
are rare near the Sun compared to 1 au,
but including events with high {\alf}icity gives a comparable number.
The original detection algorithm excluded {\alf}ic events because the GS equation is only valid for magnetostatic structures with no velocity fluctuations (Appendix~\ref{sec:theory})
and because torsional {\alf} waves have a similar observational signature to flux ropes, but with high {\alf}icity \citep{marubashi_torsional_2010,yu_small_2016,higginson_structured_2018}.
However, there have also been observations of {\alf} waves inside SMFRs
\citep{gosling_torsional_2010,shi_parker_2021}. In fact, from a theoretical standpoint, there is every reason to expect torsional {\alf} waves to form within flux ropes due to various disturbances (see \citet{gosling_torsional_2010} and references therein).
Nevertheless, previous 1 au studies based on small event databases did not contain many
events with field-aligned flows \citep{gosling_torsional_2010}
and only a small percentage of event candidates were excluded due to field-aligned flows in the GS-based original catalog.

We introduce an improved version of the GS-based automated detection algorithm.
Our implementation has the following improvements over the original algorithm:
(1) we significantly improved the computational efficiency so that supercomputer resources
are unneeded and the algorithm can be applied to higher resolution data (we used a consumer-level computer to process 20x higher resolution data than the data processed by supercomputer clusters to generate the original catalog); 
(2) we incorporated the full GS reconstruction
into the detection algorithm to provide more information about the SMFRs and to eliminate the need for the threshold on $\avg{B}$ to eliminate small fluctuations;
(3) we used the generalized GS equation for the case of a field-aligned flow with a constant {\wal} slope (as also done by \citet{chen_small-scale_2021,chen_small-scale_2022} for PSP, but not previously done for \emph{Wind}).
Beyond the new algorithm,
we present new findings on the statistical properties of SMFRs.
The validity of the statistical findings is strengthened by the improved reliability of the new algorithm, but they are based on a revised analysis, not on the improved reliability of the algorithm.

This paper is structured as follows.
Section~\ref{sec:implementation} describes our improved detection algorithm,
Section~\ref{sec:benchmark} benchmarks the algorithm against simulated measurements,
Section~\ref{sec:applicationtowind} describes the data the algorithm was applied to,
Section~\ref{sec:spatialtemporal}
analyzes the size and occurrence of SMFRs,
Section~\ref{sec:properties} analyzes physical properties of the SMFRs,
and Section~\ref{sec:discussion} contains discussion and conclusions.

\section{Improved Detection Algorithm} \label{sec:implementation}

\subsection{Motivation}

The improvements to the algorithm are motivated by a need for
increased performance
and the need for a better criteria to distinguish small fluctuations
and {\alf} waves from SMFRs, discussed below.

\subsubsection{Performance}

The original detection algorithm
is limited by its exhaustive-search nature. That is, it iterates over every possible interval, 
trying every possible axial orientation.
The whole process is then repeated with the smaller sliding window lengths.
Even coarse spacing between trial axes,
over a hundred trial axes must be used and most calculations must be repeated for each orientation.
This is not only inconvenient, but limits the scientific application of the algorithm.
The distribution of SMFR durations found in the original catalog continues to increase asymptotically down to the smallest window length used (approximately 10 minutes).
Since 21 years of 1-minute cadence data required days of supercomputer time
to process, applying the original algorithm to higher cadence data to detect smaller SMFRs would be
computationally prohibitive.

\subsubsection{Elimination of Small Fluctuations and the Magnetic Field Strength Threshold}

\begin{figure}
    \centering
    \includegraphics[width=.5\textwidth]{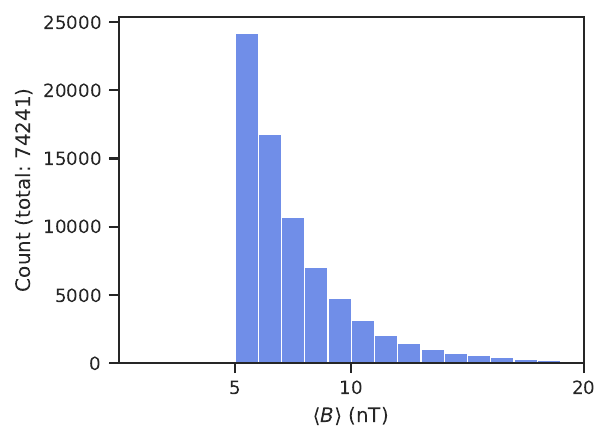}
    \caption{Distribution of $\avg{B}$ (the magnetic field strength averaged over each flux rope's interval) in the original catalog.} 
    \label{fig:original_catalog_Bdistr}
\end{figure}

Since SMFRs were originally considered
to be transient structures with elevated magnetic field strength $B$,
and the average IMF $B$ is 5 nT,
the original algorithm required that $B > \SI{5}{\nano\tesla}$.
However, the resulting distribution of $B$ in SMFRs in the original catalog is cut off right at
the peak (Figure~\ref{fig:original_catalog_Bdistr}).
Therefore, it is likely that ${\sim}$50\% of the SMFRs are excluded,
which one can expect to cause significant statistical bias.
Indeed, as mentioned in the introduction, we have
previously demonstrated that the physical properties
of SMFRs are mostly the same as the background solar wind other than
the fact that $B > \SI{5}{\nano\tesla}$.

\begin{figure}
    \centering
    \includegraphics[width=.8\textwidth]{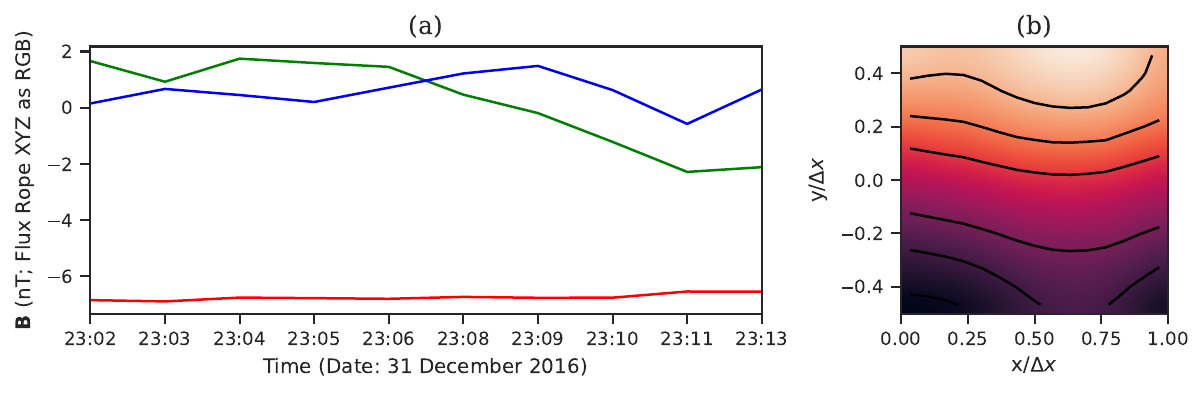}
    \caption{Magnetic field measurements in the flux rope coordinate system (a), as well as the GS reconstruction (b), of event \#74238 from the original catalog. In (a), the red, green, and blue lines correspond to $B_x$, $B_y$, and $B_z$, respectively. In (b), the brightness represents $A$. For a flux rope, we expect that $B_y$ should exhibit bipolarity and $B_z$ should increase towards the center, consistent with the signature observed in (a). Despite that, the GS reconstruction in (b) does not contain any closed transverse field lines.}
    \label{fig:original_catalog_reconstruction_example}
\end{figure}

Is requiring $B > \SI{5}{\nano\tesla}$
an effective method to exclude small fluctuations?
Despite this threshold, it appears that the original catalog contains many events that appear to simply be a small fluctuation in the magnetic field direction.
We find that an objective way to distinguish small fluctuations from SMFRs is to perform the full GS reconstruction on a given interval to test if it really contains a flux rope (closed transverse field lines in 2D).
Figure~\ref{fig:original_catalog_reconstruction_example} shows an example. While Figure~\ref{fig:original_catalog_reconstruction_example} (a) exhibits a magnetic field signature consistent with the crossing of a flux rope at a high impact parameter, the reconstruction in Figure~\ref{fig:original_catalog_reconstruction_example} (b) does not confirm the existence of any closed transverse field lines. It is possible that the event is a flux rope that the spacecraft passed through far from its center, but it could just be a small kink in the magnetic field. The original detection algorithm only tests the hypothesis that each transverse field line is crossed twice. We should also check whether any of those transverse field lines are closed.
Using GPUs, it is not computationally prohibitive to perform the GS reconstruction of many
sliding windows. As Figure~\ref{fig:original_catalog_reconstruction_example} shows, this can eliminate many false positives and thus make the results more reliable.

\begin{figure}
    \centering
    \includegraphics[width=.9\textwidth]{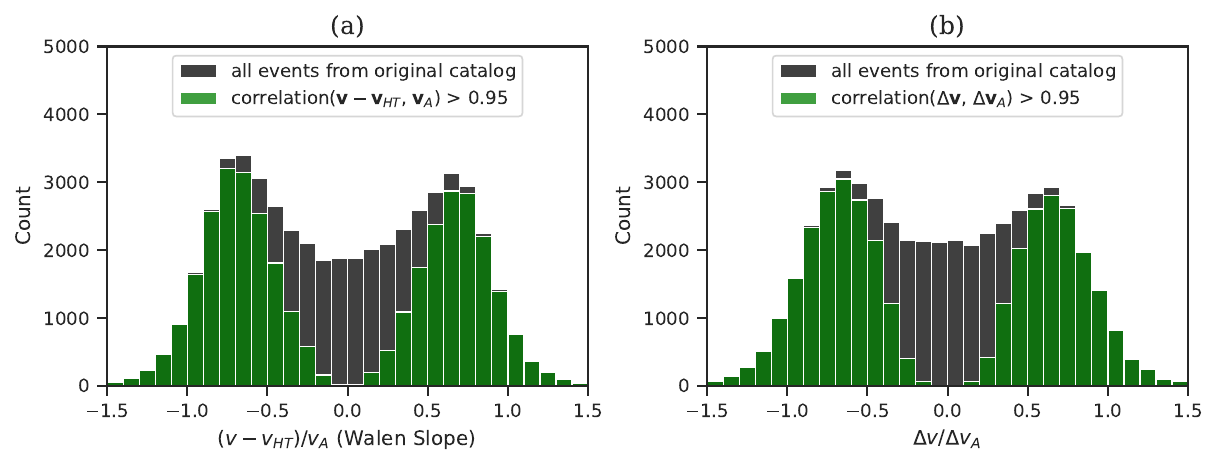}
    \caption{(a) Distribution of {\wal} slope calculated using $\vHT$ for the events original catalog.
    The histogram is cut off at $\pm1.5$. Note that for most events, the {\wal} slope $\abs{R_w} > 0.3$, in contrast to the requirement the original detection algorithm that all events have $\abs{R_w} < 0.3$. The green distribution is the subset where the linear correlation is strong, showing that for the events with a well-determined slope, the {\wal} slope follows a normal distribution. (b) Same as (a) except using the reference frame agnostic method to determine {\alf}icity as introduced by \citet{chao_walen_2014}. In this case, the slope is the relationship between $\dv*{\vb{v}}{t}$ and $\dv*{\vb{v}_A}{t}$, which does not depend on the reference frame.}
    \label{fig:original_catalog_walen_distr}
\end{figure}

\subsubsection{{\alf}icity}

We have discovered that the apparent difference in abundance of {\alf}ic events
between PSP and 1 au is primarily due to a difference in methodology,
not a physical difference.
The {\wal} slope measuring {\alf}icity must be found in the zero electric field frame of reference $\vHT$ (Appendix~\ref{sec:theory}), not in the average velocity frame.
But prior to PSP applications, the flux rope velocity $\vFR$ was approximated as $\avg{\vb{v}}$
instead of $\vHT$ due to the large volume of data and the computational inefficiency of the original detection algorithm.
This has a significant effect on the calculation of the {\wal} slope, since the component-by-component {\wal} slope must be calculated in the HT frame \citep{khrabrov_dehoffmann-teller_1998,paschmann_proper_2008}.

The difference between using $\avg{v}$ and $\vHT$ can be seen as follows: Suppose that the total bulk velocity is the background velocity plus a fluctuation $\vb{v}_0 + \delta \vb{v}$, and that the fluctuation is a field-aligned flow such that it can be written in terms of the {\alf} velocity vector as $\delta \vb{v} = R_w \vb{v_A}$, where the constant of proportionality $R_w$ is the {\wal} slope. If one attempts to evaluate $R_w$ using the average velocity frame instead of the HT frame by estimating $\delta \vb{v} = \vb{v} - \avg{v}$, an issue arises: this approximation is only valid if $\avg{\vb{v_A}} = 0$. In a magnetic flux rope, $\avg{\vb{v_A}}$ may have a nonzero component in the $\vu{x}$ direction and always has a significant nonzero component in the $\vu{z}$ direction (Figure~\ref{fig:coords}). Therefore, the different components will be misaligned, making it impossible to calculate the {\wal} slope, in fact resulting in an estimated {\wal} slope near zero even for an event with significant field aligned flows.

We calculated the {\wal} slope for each event in the original catalog using $\vHT$ instead of $\avg{v}$ using the \emph{Wind} \citep{wilson_quarter_2021}
measurements of magnetic field (via the MFI instrument; \citet{lepping_wind_1995}) and plasma moments (via the SWE instruments; \citet{ogilvie_swe_1995}).
Figure~\ref{fig:original_catalog_walen_distr} (a) shows the distribution of the {\wal} slope for the events in the original catalog. Even though the original algorithm excludes events with $\abs{R_w} > 0.3$, less than 25\% of the events actually meet that threshold when $\vHT$ is used to calculate the {\wal} slope.

One might wonder if $\vHT$ artificially introduces
the field-aligned flow by minimizing $\vb{v} \cross \vb{B}$.
Figure~\ref{fig:original_catalog_walen_distr} (b) shows the same result using a reference-frame independent method to determine {\alf}icity \citep{chao_walen_2014}: instead of taking the slope of $\delta \vb{v} \propto \vb{v}_A$,
one can take the derivative of both sides and then the slope can be evaluated in any reference frame.
The result using the frame-independent method is the same as the result obtained using the HT frame: most events in the original catalog at 1 au are highly {\alf}ic.

\subsection{Optimization}

An easy way to optimize the detection algorithm is to take advantage of the fact that the
same computations are applied to each interval.
In Python, loops are slow. This problem can be avoided by processing multiple intervals simultaneously via matrix computations.
On a CPU, this benefit is a consequence of the limitation of the Python language.
However, GPUs can take greater advantage of this, because unlike CPUs, they apply a single instruction to
large arrays in parallel. In our implementation, we utilize batch processing
and run our code on a GPU using the PyTorch software package for matrix computations.
The interpolation operation necessary for the computation of $R_\mathrm{diff}$ is
made possible by using another software package (\url{https://github.com/aliutkus/torchinterp1d}).
The use of GPU processing allows our implementation to process large volumes of data on a consumer-grade computer instead of a supercomputer.

We further improved the performance of the algorithm by reducing the search space for axial orientation
$\vu{z}$ (Figure~\ref{fig:coords}).
This is possible due to the following realization.
For an acceptable flux rope interval, the same field line is observed at the beginning and end.
Since the poloidal flux function $A$ is a field line invariant, the final $A$ equals the last $A$, that is, $A_f = A_0$.
Since the difference in $A$ is given by integrating $B_y$ (Appendix~\ref{sec:theory}), the integral of $B_y$ over the interval must then be zero. Therefore, the average magnetic field vector must be perpendicular to $\vu{y}$, since $\avg{\vb{B}}\vdot \vu{y} = \avg{\vb{B}\vdot \vu{y}} = -\abs{v_x} \int_{t_0}^{t_0+\Delta t} B_y dt / \Delta t= 0$ (where the average $\avg{\vb{B}}$ must be calculated as the integral of each component over the interval divided by the length, rather than the mean of each component).

So $\expval{\vb{B}}$ is perpendicular to $\vu{y}$, in addition to $\vFR$ (following from the definition of the vertical direction $\vu{y}$; Figure~\ref{fig:coords}). Then, even before knowing $\vu{z}$, $\pm\vu{y}$ is already determined. This explains why the uncertainty in $\vu{z}$, often referred to as azimuthal uncertainty when viewed in the GSE coordinate system, is typically much greater than the uncertainty in $\vu{y}$, as observed by \citet{hu_reconstruction_2002}. Once $\pm \vu{y}$ is known, $A$ is known up to a constant factor of $\pm v_x$. However, for the purpose of finding $\vu{z}$, only the relative values of $A$ are needed to specify the field line corresponding to each measurement, so we can simply normalize $A$ to start at $0$ and peak at $1$.

To find $\vu{z}$, we still need to minimize the difference residue $R_\mathrm{diff}(\theta)$ (Appendix~\ref{sec:theory}) where $\theta$ is the angle between $\vu{z}$ and $\vFR$ about $\vu{y}$. The only component of $P_t$ affected by the choice of
$\theta$ is $B_z^2/2\mu_0$, so we only need to minimize the contribution from the $B_z$ term
(and $B_z$ itself is a field line invariant; Appendix~\ref{sec:theory}).
Originally, we would have to calculate $B_z(t) = \vb{B}(t) \vdot \vu{z}$ for
each $\vu{z}$, then linearly interpolate the values before and after the peak $A$ onto $A(t)$,
yielding $B_z^{(1)}(A)$ and $B_z^{(2)}(A)$. Finally, we would calculate $R_\mathrm{diff} = \sqrt{\avg{(B_z^{(2)}(A) - B_z^{(1)}(A))^2}}/(\max(B_z) - \min(B_z))$ for each $\theta$.
However, linear interpolation is a fairly expensive operation. Instead of performing it each time,
we can take advantage of the fact that the normalized $A$ is fixed and is not affected by $\theta$,
so the linear interpolation for each point is just a linear combination of the original values of $B_z$
with coefficients independent of $\theta$ or $\vu{z}$. So if instead we interpolate the vector $\vb{B}$ by interpolating its components, we can construct an $N\times3$ matrix of difference vectors $\vb{B}^{(2)}(A) - \vb{B}^{(1)}(A)$, where $N$ is the number of measurements.
If we multiply it by the $3\times M$ matrix of $M$ possible $\vu{z}$ orientations,
each element of the resulting $N\times M$ matrix is $B_z^{(2)}(A) - B_z^{(1)}(A)$ for the
measurement corresponding to the row and the possible $\vu{z}$ corresponding to the column.
Squaring the values of the matrix and summing over the rows yields the numerators for calculating $R_\mathrm{diff}(\theta)$. From there, the same procedure can be applied to the original $N\times3$ matrix of measured magnetic field vectors $\vb{B}$ to obtain $B_z(t)$ for each orientation,
which can then be used to find $\max(B_z) - \min(B_z)$ and thus the denominator of $R_\mathrm{diff}$.

The procedure outlined above makes the assumption that the interval is perfectly selected and does not need to be trimmed further.
Making the first assumption requires the use of very narrowly spaced sliding window lengths. For example,
\citet{hu_automated_2018} used 1-minute cadence data with sliding windows 5 minutes apart in length, trimming the window to be shorter if necessary.
With this new procedure, a 1-minute separation between window lengths would be necessary due to the assumption that the interval is already trimmemd. However, this only requires visiting each flux rope candidate at most five additional times, whereas the reduction by finding $\vu{y}$ first is significantly greater.

Another assumption made is that
$\vFR \cross \avg{\vb{B}} \neq 0$. When this relationship is not satisfied, we need to resort to a full trial-and-error process. We do this by trying every possible $\vu{y}$ (with 1 degree spacing) and finding the one that has a single stationary point, the absolute value of the last $A$ value being less than 10\% of the absolute value of the peak $A$, and that minimizes $R_\mathrm{diff}$ when $\vu{y}$ is not well specified. We evaluate $\vu{y}$ as being well-specified by taking the cross product of the estimated $\vu{y}$ and $\vFR$, and validating the average magnetic field along this perpendicular direction is not less than 10\% of the average of the magnetic field magnitude. The handling of this edge case is not particularly important and only introduces a minimal performance impact, since the average magnetic field direction being parallel to the velocity is rare (even near the Sun, where the magnetic field is approximately radial, the Parker Solar Probe's own motion perpendicular to the radial direction ensures that the velocity relative to the spacecraft is not purely radial).

The new procedure for finding the axial orientation provides a substational improvement in
the performance of the algorithm. Additionally, it makes it easy to increase the angular precision from more than 10 degrees to less than 1 degree without an absurd computational cost, since the search for $\vu{z}$ is now reduced to varying 1 angular parameter instead of every combination of two angular parameters. For example, if both the azimuthal and latitudinal separations were 1 degree, over 30,000 trial axes would be necessary for every interval.

\begin{figure}
    \centering
    \includegraphics[trim={8cm 0 8cm 0},clip,width=\textwidth]{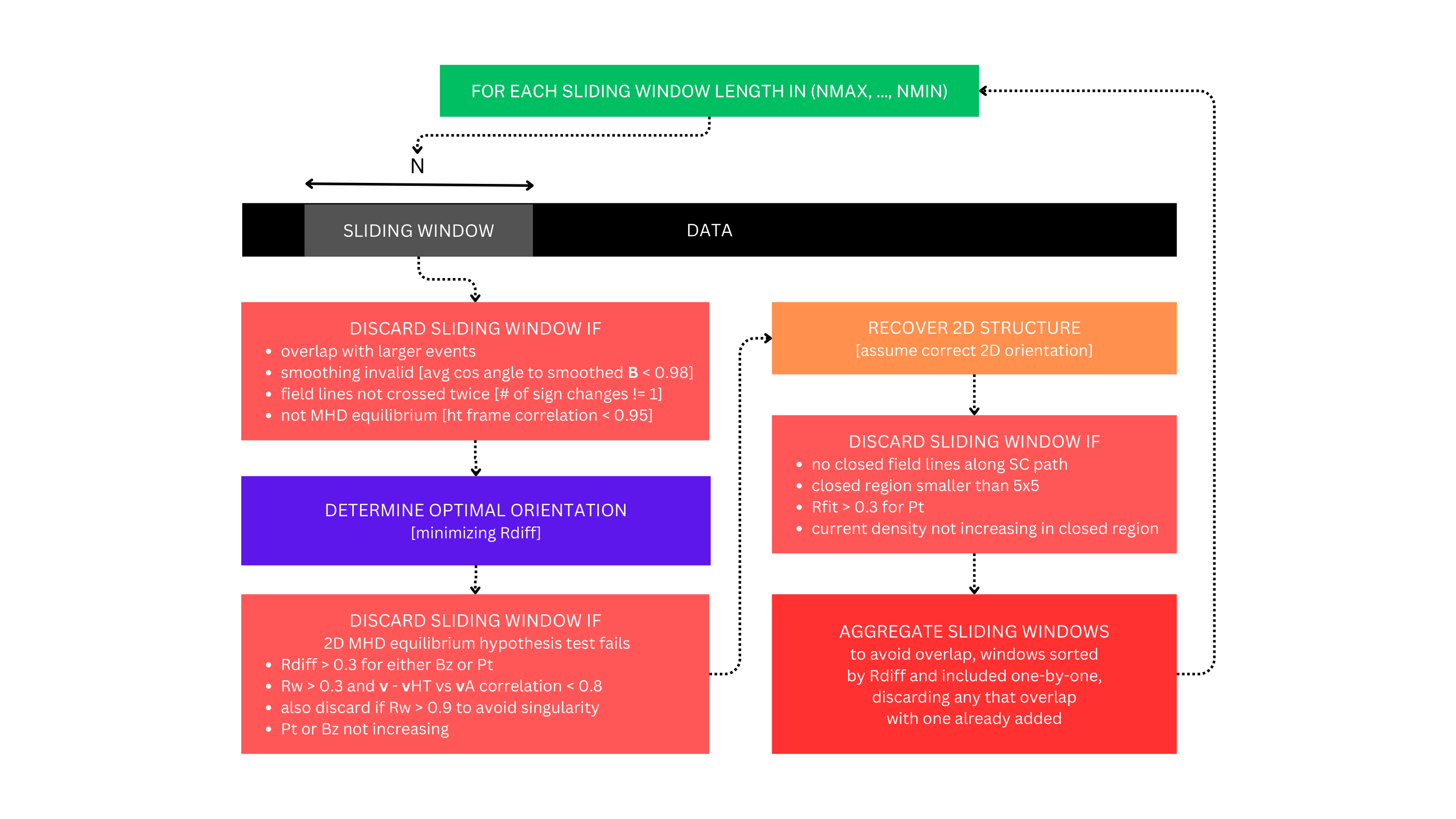}
    \caption{Flow chart describing the improved detection algorithm.}
    \label{fig:flowchart}
\end{figure}

\subsection{Algorithm}

In this section, we describe the algorithm step by step for reproducibility purposes.
The overall procedure is outlined in Figure~\ref{fig:flowchart}.
Essentially, the algorithm checks every single interval for a set of interval lengths,
tests whether the interval is compatible with a flux rope signature: (1) magnetostatic or MHD equilibrium is validated by finding and validating $\vHT$; (2) smoothness of the magnetic field in the interval at the given scale is tested by comparing the raw measurements to a smoothed version of the measurements; (3) the optimal 2D orientation is found and the hypothesis of a 2D structure where every field line is crossed twice is tested
by verifying that $A$ that $P_t' = P_t'(A)$; (4) the structure is checked for closed field lines via GS reconstruction; (5) the remaining sliding windows are filtered to avoid overlap, giving preference to larger event candidates over smaller ones.

\subsubsection{Selecting Smooth Intervals Exhibiting MHD Equilibrium}

First, we process the data
using a set of sliding windows. Our goal is to use window lengths that span multiple orders of magnitude, and GPUs have limited memory, so some additional steps are required. We set a maximum processing resolution $N_\mathrm{max}$. If the sliding window length $N_\mathrm{window} > N_\mathrm{max}$,
then we must downsample the data so that the windows are not prohibitively memory intensive. For example, using 3 second cadence data, a window length approximately equal to 3 days would contain approximately $10^5$ data points. We find that $N_\mathrm{max} = 64$
does not significantly affect the results compared to higher resolutions despite providing a substantial performance gain. 
To downsample the data, we first perform a boxcar average with both length and step size set to $\lfloor N/N_\mathrm{max} \rfloor$, so that the remaining downsampling factor is less than a factor of 2. Then, we use linear interpolation to resample the data so that the window length becomes $N = N_\mathrm{max}$. If $N < N_\mathrm{max}$, then the window length is left as is.
To ensure that the event's magnetic field fluctuation is ``smooth'' so that the downsampling and smoothing employed throughout the algorithms is valid,
we require that
the average cosine angle
between the downsampled vectors
and their moving average with kernel size $\lfloor N/10 \rfloor$ (rounded up to the nearest odd number if even)
be at least 0.98.

Next, we evaluate $\vHT$ and the average $\vb{B}$ (using the trapezoid rule since the same is used for evaluating the poloidal flux function $A$) for each sliding window.
Using these quantities, we evaluate the vertical direction $\vu{y}$.
The correlation coefficient for $\vHT$ \citep{khrabrov_dehoffmann-teller_1998} is required to be at least 0.95, and
$B_y$ is required to have exactly one change of sign
after (applying the same filter used for the smoothness check
to avoid excluding events containing embedded fluctuations).

\subsubsection{Finding 2D Orientation and Validation of 2D Hypothesis}

For the sliding windows that remain,
we evaluate the minimum residue $\vu{z}$ with 256 trial axes centered around the direction of the average magnetic field uniformly spread between $\pm\SI{90}{\degree}$ about the estimated $\vu{y}$.
The selected $\vu{z}$ is guaranteed to result in a trimmed $A$, as it is perpendicular to the already estimated $\vu{y}$. The resolution n of the distinction between $\vu{x}$ and $\vu{z}$ is $\SI{180}{\degree}/256 \approx \SI{0.7}{\degree}$, but our benchmarking suggests that the actual uncertainty is on the order of \SI{10}{\degree}.
We require $R_\mathrm{diff} < 0.3$ (Appendix~\ref{sec:theory}) separately for $B_z$ and $P_t$. Previous GS-based detection studies used a stricter threshold for $R_\mathrm{diff}$ since they did not directly solve the GS equation to validate the flux rope structure and relied primarily on a low $R_\mathrm{diff}$ for detection. However, many events studied using GS reconstruction in the literature had significantly higher values of $R_\mathrm{diff}$, and even flux ropes in MHD simulations can have higher $R_\mathrm{diff}$. We find that in conjunction with our validation through GS reconstruction, a threshold of 0.3 provides satisfactory results. Note that we do not use the extra factor of $1/\sqrt{2}$ in our definition of $R_\mathrm{diff}$, which would make our threshold just over 0.2 by the definition used in \citet{hu_automated_2018}. 

Since it seems that most SMFRs have non-negligible {\alf}icity
(Figure~\ref{fig:original_catalog_walen_distr}), we must account for the {\alf}icity when calculating the generalized transverse pressure $P_t$. We use the {\wal} slope to estimate a constant {\alf} Mach number in the flux rope frame of reference $M_A$, so that we may use Equation~\ref{eq:modifiedGS}
to calculate the generalized $P_t$.
When the {\wal} slope is greater than 0.3, we require that the correlation coefficient between $\vb{v} - \vHT$ and $\vb{v}_A$ be at least 0.8, and we exclude events with a {\wal} slope greater than 0.9 to avoid a singularity in Equation~\ref{eq:modifiedGS} \citep{chen_small-scale_2021,chen_small-scale_2022}.

\subsubsection{Full GS Reconstruction}

We perform the full GS reconstruction
for all of the sliding windows that pass the above tests.
\citep{sonnerup_grad-shafranov_2006,teh_gradshafranov_2018}.
The settings must work well when applied to a large number
of events automatically without any manual adjustments. This is especially important
due to the sensitivity of solving the GS equation as an initial value problem.
Based on our experimentation, we find that it is suitable to use a third-order polynomial to fit
$P_t(A)$ and take its derivative to derive the current density used for the reconstruction.
The validity of the polynomial fit is ensured by requiring the events to have $R_\mathrm{fit}$ no more than 0.3.
Additionally, we require that the axial current density increases towards the center of the closed region,
since real flux ropes carry strong axial currents that are strongest
at their center.
We added this requirement because SMFRs should carry strong axial currents to generate their poloidal magnetic field,
and because in the MHD simulation we benchmarked
the algorithm on, all events where the reconstruction
did have a monotonically increasing current density
were false positives.

The size of the reconstruction domain is fixed at 11x11 pixels with step size $\Delta y = 0.1\Delta x$,
and we use the standard three-point smoothing introduced in \citep{hau_two-dimensional_1999}
for the stability of the numerical integration. The 1D measurements are downsampled to 11 data points and the reconstructed cross section is 11x11 pixels (5 pixels above and below the spacecraft path). For values of $A$ outside of the measured range, the lower tail of $P_t(A)$ is extrapolated using a decaying exponential so that the current density decays outside of the measured field lines, while for the upper tail, it is simply allowed to continue according to the polynomial fit.
Once the reconstructed map is obtained, events without a core with closed transverse field lines according to the procedure previously mentioned are excluded. We also require that the closed region be more than 4 pixels wide and 4 pixels tall and that it overlap with the strip measured by the spacecraft.
The procedure for finding the core region is described in Appendix~\ref{sec:core_region_algo}.

\subsubsection{Cleanup of Overlapping Windows}

Once the candidate windows are obtained for a given sliding window size, we follow the same procedure as in the original algorithm: For a given window length, we use the greedy algorithm to remove overlapping events. However, rather than prioritizing events by their end time, we prioritize them by $R_\mathrm{diff}$. The gaps between larger events are filled with events detected from smaller sliding window sizes, but larger events are prioritized over smaller events.

\section{Benchmarking Against Simulated Measurements} \label{sec:benchmark}

\subsection{Small Fluctuations}

\begin{figure}
    \centering
    \includegraphics[scale=.55]{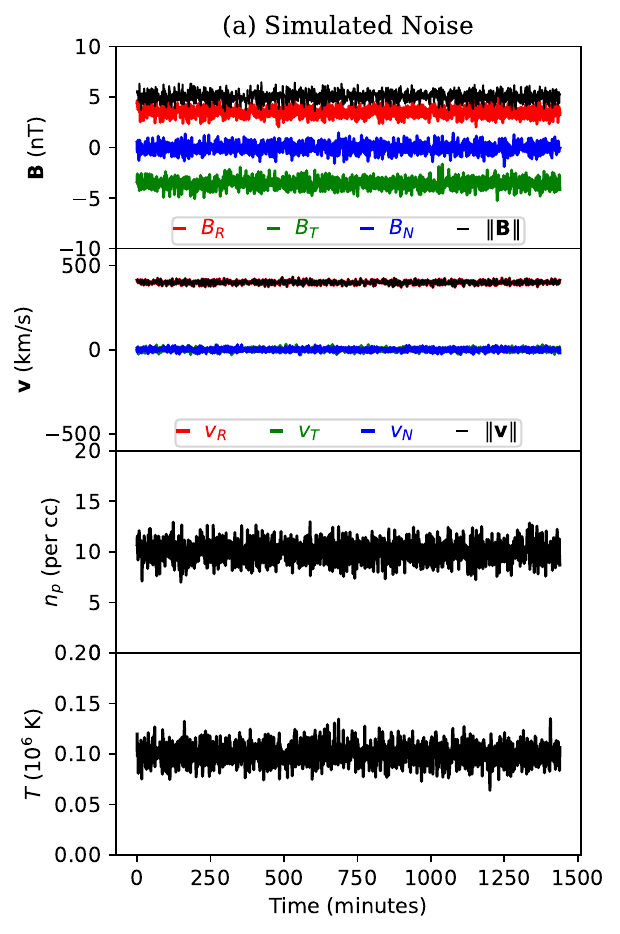}
    \includegraphics[scale=.55]{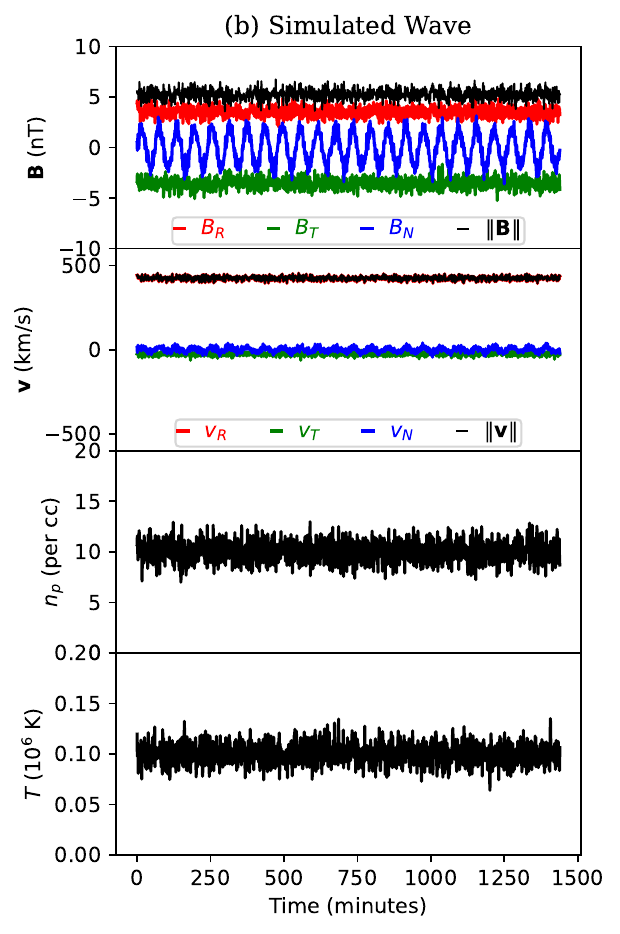}
    \includegraphics[scale=.55]{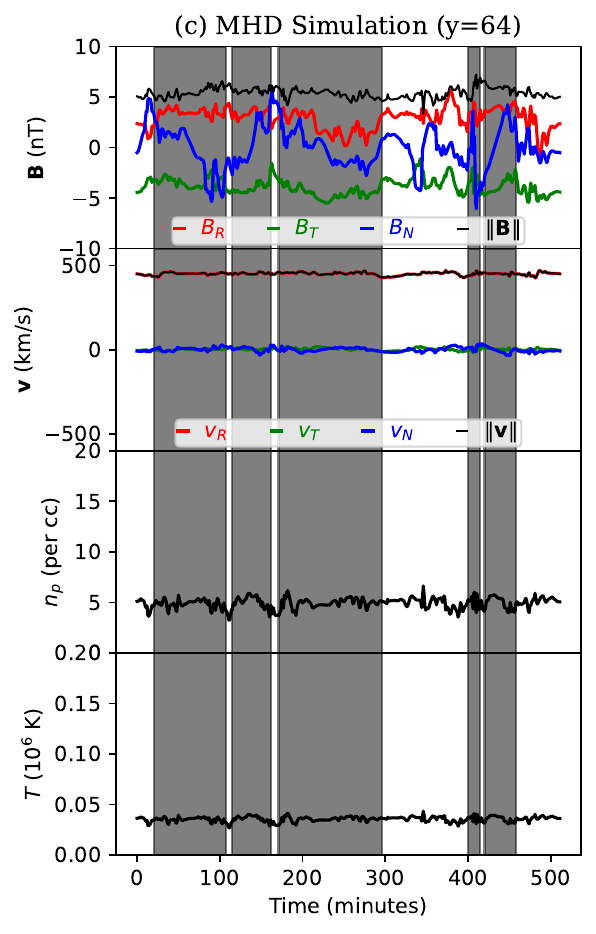}
    \caption{Examples of the results on simulated timeseries. Intervals detected as flux ropes are shaded. The columns are: (a) background field plus noise (b) background field plus pure {\alf} wave (c) MHD simulation}
    \label{fig:synthetic}
\end{figure}

Since the number of events detected is so large
and most of them are quite short, one could reasonably be concerned that a significant portion of the detected events are not flux ropes at all, but rather small fluctuations, such as noise or waves.
Here, we demonstrate that such fluctuations are not a significant source of false positives.

First, we generate simulated, noisy data. All vector quantities here are in radial-tangential-normal (RTN) coordinates. We generate 1 day of 1 minute cadence simulated data. The magnetic field data is a constant Parker spiral aligned field plus Gaussian noise $\boldsymbol{\mathcal{N}}$ with mean $\mu$ and standard deviation $\sigma$: $\vb{B} = \SI{5}{\nano\tesla} (\sqrt{2}\vu{R} - \sqrt{2}\vu{T}) + \boldsymbol{\mathcal{N}}(\mu=0, \sigma=\SI{0.5}{\nano\tesla})$. The proton number density $n_p = \mathcal{N}(\mu=\SI{10}{\centi\meter^{-3}}, \sigma=\SI{1}{\per\centi\meter^{-3}})$,
the proton temperature $T = \mathcal{N}(\mu=\SI{1e5}{\kelvin}, \sigma=\SI{1e4}{\kelvin})$,
and the proton bulk velocity is radial plus noise $v = \SI{400}{\kilo\meter\per\second} \vu{R} + \boldsymbol{\mathcal{N}}(\mu=0, \sigma=\SI{10}{\kilo\meter\per\second})$. The data is plotted in Figure~\ref{fig:synthetic} (a). Additionally, we generate a similar artificial data interval with a simple {\alf} wave added: $\vb{B} \to \vb{B} + \sin((2\pi/\SI{60}{\minute}) t) \vu{N}$ and $\vb{v} \to \vb{v} + \vb{v}_A$. The data with the wave is plotted in Figure~\ref{fig:synthetic} (b).

As expected, neither the noisy simulated data nor the wavy simulated data had any flux ropes detected with windows ranging from 10 to 360 minutes. Though we did not reproduce it here, after many repetitions with a higher $\sigma$ for the magnetic field noise, we managed to get one single event of only 10 data points to be detected. However, that is far too rare to explain the large number of SMFRs that are detected in a given day of real data. Also not shown here, we have verified that even with the velocity fluctuations in the simulated data removed or with a magnitude lower than the {\alf} speed, the simulated wave is not detected as an event by our algorithm. Thus,
noise and pure transverse {\alf} waves are not likely to be a significant source of false positives.

\subsection{MHD Simulation}

To validate the applicability of our method to realistic SMFRs, we
tested our new algorithm on SMFRs generated from a 2.5D MHD turbulence simulation.

\begin{figure}
    \centering
    \includegraphics[width=0.8\textwidth]{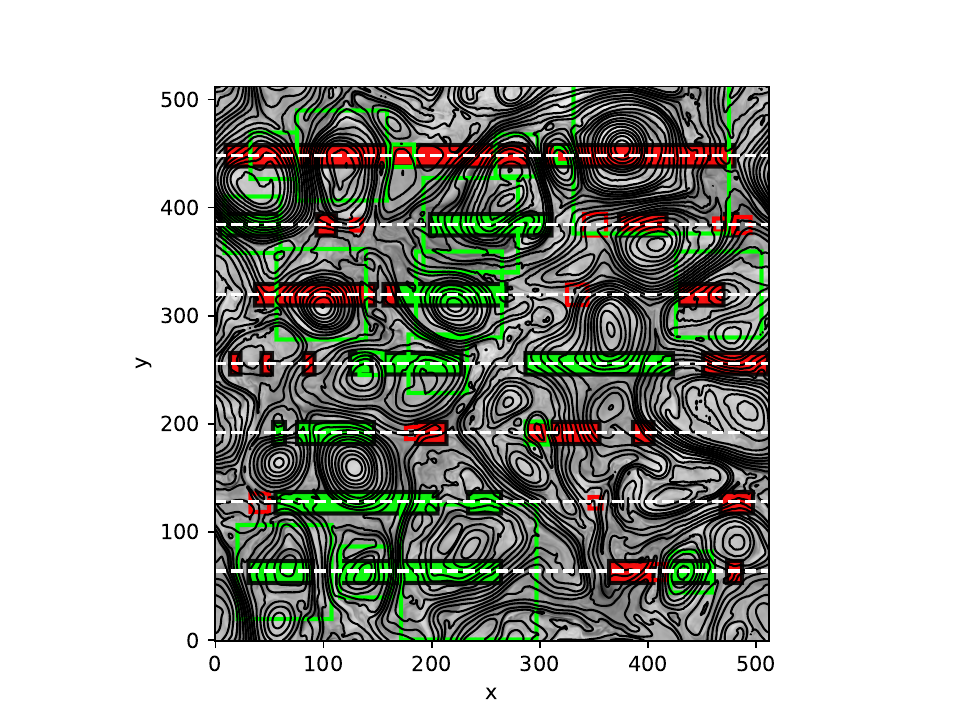}
    \caption{Results from applying the improved algorithm to a 2.5D MHD turbulence simulation.
    The brightness represents the strength of $B_z$ and the blue streamlines represent the transverse field lines. 
    The dashed lines are virtual spacecraft paths through the simulation. The square boxes represent sliding window intervals that were detected as flux ropes by the new algorithm.
    The rectangular boxes represent the intervals detected as events by the original algorithm. Green boxes are classified as true positives, whereas red boxes are classified as false positives.}
    \label{fig:benchmark_mhd}
\end{figure}

We test our algorithm in numerical simulations of compressible  MHD (CMHD). The CMHD equations are solved in a $2.5$D square box of length $2\pi L_0$ in either direction, with 4096 points per side, using a pseudo-spectral code as described in Ref.~\citep{vasconez_kinetic_2015,perri_numerical_2017,pecora_identification_2021}.
The simulation was performed in the $x$-$y$ plane with a mean magnetic field $B_0=1$ along the $z$ direction.  Velocity and magnetic field fluctuations have all three Cartesian components. The algorithm is stabilized by fourth-order hyperviscosity that suppresses very small-scale, numerical effects. The parameters of the simulation are appropriate to describe solar wind conditions, magnetic fluctuations are such that $\delta b/B_0=1/2$ and plasma $\beta\sim0.5$, with $\delta b$ total r.m.s magnetic fluctuation amplitude and $\beta$ ratio of kinetic to magnetic pressures. 
The initial fluctuations are chosen with random phases, for both magnetic and velocity fields, in a shell of Fourier modes with $3\leq |{\boldsymbol{k}}|\leq 5$. The decaying CMHD simulation quickly develops turbulence and small-scale dissipative structures. The magnetic field power spectrum (not shown here) manifests a typical scaling $P(k)\propto k^{-5/3}$. The turbulent pattern is represented in Figure~\ref{fig:benchmark_mhd}.

\begin{deluxetable}{lrrrrrr}
\label{tab:comparison}
\tablecaption{Quantitative comparison between both algorithms when applied to the MHD simulation and with or without events below 30 data points.}
\tablehead{\colhead{Algorithm} & \colhead{Time per row} & \colhead{True Positive (TP)} & \colhead{TP ($N > 30$)} & \colhead{Good Reconstruction (Rec)} & \colhead{Rec ($N > 30$)}}
\startdata
Original & ${\sim}4$m & 14 (42\%) & 12 (57\%) & N/A & N/A \\
New & ${\sim}2$s & 18 (64\%) & 14 (100\%) & 35\% & 79\% \\
\enddata
\end{deluxetable}

Figure~\ref{fig:benchmark_mhd} shows the results of applying the detection algorithm to the MHD simulation (downsampled to 512x512).
Virtual spacecraft were sent through several paths to take virtual measurements,
to which the detection algorithm was applied.
An example for a single spacecraft path is shown in Figure~\ref{fig:synthetic} (c).
Also included are rectangles representing
events detected using the original algorithm.
We automatically classify a detected event as a true positive if (1) the interval has a
closed field line and (2) the true $B_y$ has only one inflection point (after smoothing).
True positives are labeled green while false positives
are labeled red.

Although the original algorithm occasionally picks up good events
that the new algorithm misses,
in many cases it either incorrectly combines multiple events into one,
or it detects some fluctuations at the boundary or within a flux rope.
In contrast, the new algorithm usually identifies the flux rope correctly,
even if it does not have an accurate reconstruction.
However, for the smallest window sizes with less than 30 data points,
even the new algorithm picks up some none-flux rope fluctuations as SMFRs.
Quantitative comparison is provided in Table~\ref{tab:comparison}.
From this comparison, it is clear that the new algorithm
is both significantly faster than the original algorithm
and significantly more reliable.

\begin{figure}
    \centering
    \includegraphics[width=.4\textwidth]{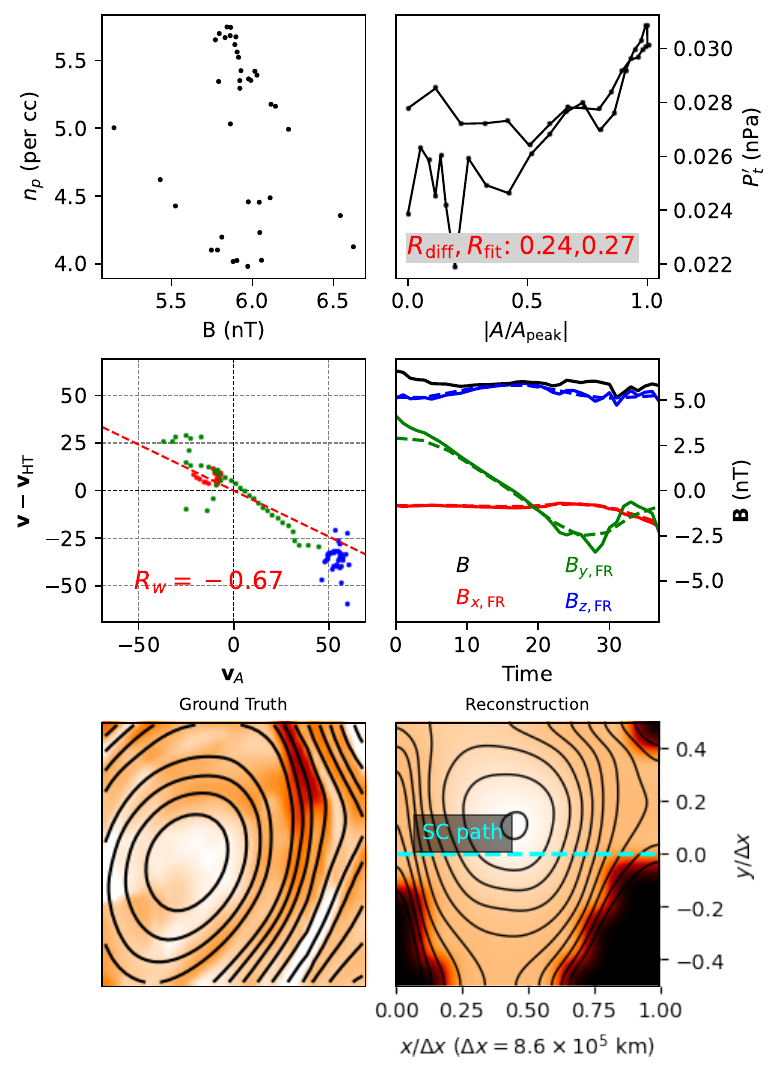}
    \includegraphics[width=.4\textwidth]{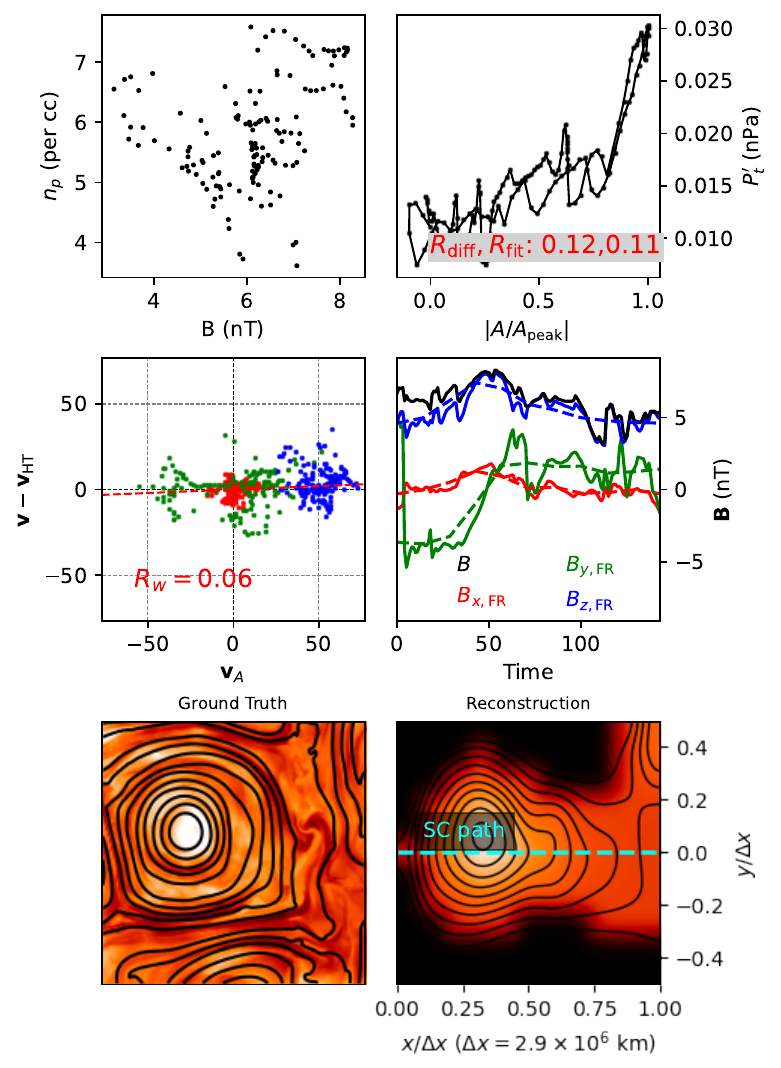}
    \includegraphics[width=.4\textwidth]{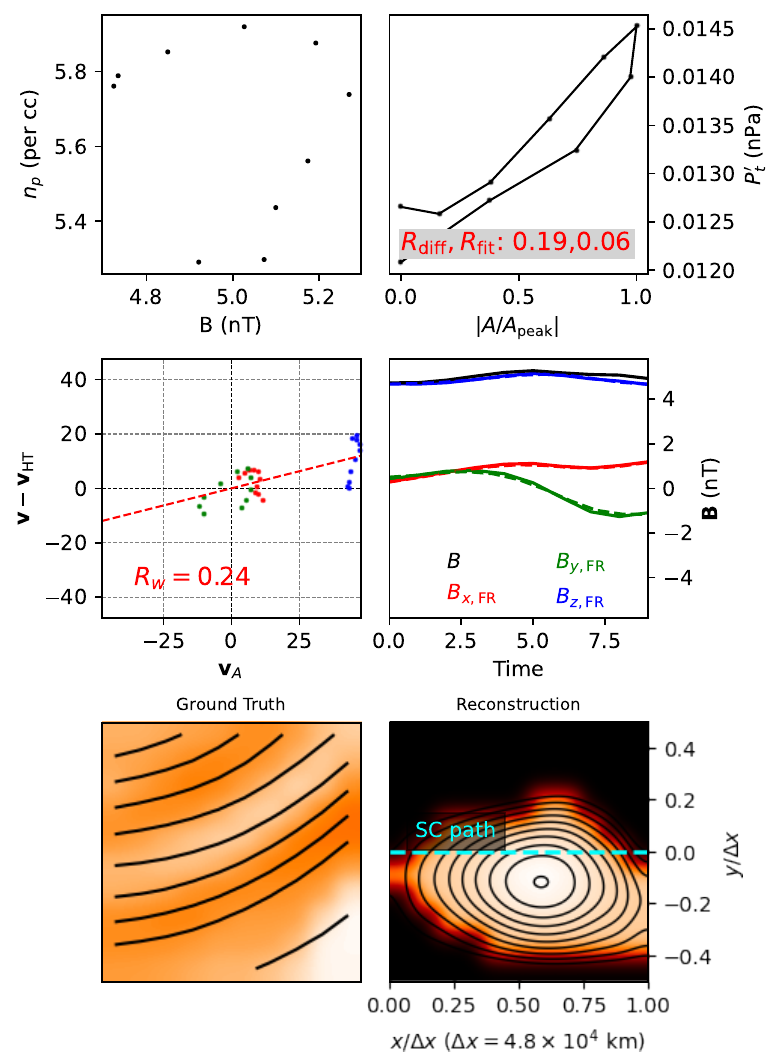}
    \includegraphics[width=.4\textwidth]{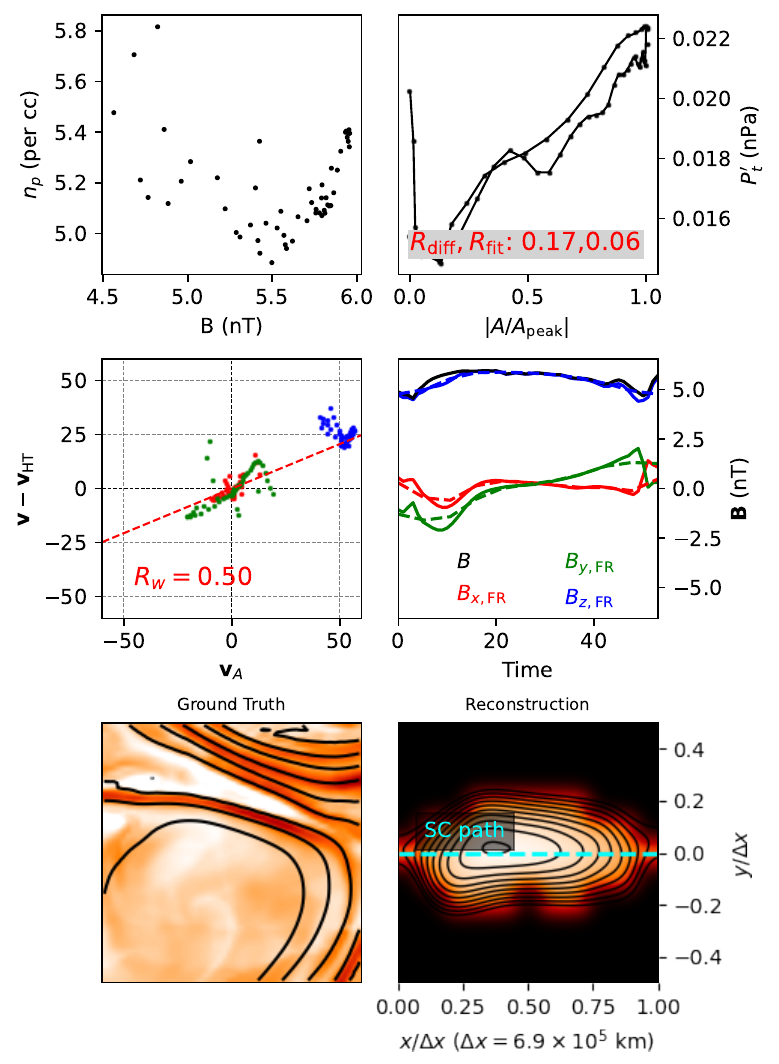}
    \caption{Four events detected by virtual spacecraft from the MHD simulation. 
    Top left: Relation between $B$ and $n_p$.
    Top right: $P_t'$ as a function of $A$ (Appendix~\ref{sec:theory}).
    Center left: Relationship between {\alf} velocity (as a vector) and the velocity fluctuations.
    Center right: Measured magnetic field components and strength (solid lines) and model magnetic field components (dashed lines) in the flux rope coordinate system.
    Bottom: Reconstructed cross section with contours at equally spaced values of the flux function $A$ (in the estimated flux rope coordinate system), along with the ground truth magnetic field (in the simulation coordinate system, not the estimated flux rope coordinate system).}
    \label{fig:sim_examples}
\end{figure}

Manual comparison between the reconstructed cross sections
and the true magnetic field geometry
suggests that the algorithm
reliably detects the presence of flux ropes (86\% of the time). However,
the reconstruction is only reliable 35\% of the time.
This appears to be because when the spacecraft crosses the flux rope
close to the edge, the wrong boundaries are selected,
so the reconstruction is inaccurate even though there is
excellent agreement with the spacecraft measurements over the 1D path that it crossed.
It is not possible to consistently distinguish which reconstructions are reliable because even the
cases with an accurate reconstruction can exhibit a relatively high $R_\mathrm{diff}$,
whereas the reconstructions that are way off can sometimes have a low $R_\mathrm{diff}$.
Therefore, for additional validation of the reconstruction, it would be valuable to incorporate
multi-spacecraft analysis in future studies. For the purposes of this study,
due to the focus on the long-term trend, we must work within
the limitation of single spacecraft measurements.
Even when the reconstructed geometry is inaccurate, the order of magnitude
of estimated parameters such as size and current density are usually reliable.

It is worth noting that, at least for this simulation, the large events (greater than 30 data points) tend to be reliable,
wheras the small events (less than 30 data points) tend to be unreliable.
However, this may be due in part to the scale of the flux ropes in the simulation being larger to begin with.
This is because it is easier to randomly satisfy a false hypothesis with a small number of data points, but events with many data points can
usually be more reliably validated.

In Figure~\ref{fig:sim_examples}, we show the detection algorithm's output for four example events based on the simulated data. The top two events are ``good'' results, whereas the bottom two are ``bad'' results. In all four cases, the algorithm correctly identified a flux rope, but in the ``bad''
cases, the reconstruction is inaccurate.
Unfortunately, we have not been able to find
any criteria that can consistently
evaluate the reliability of a reconstruction. For example, the bottom
events have lower $R_\mathrm{diff}$ than the top left event.
Setting a lower threshold on $R_\mathrm{diff}$ does not appear to improve the quality of the events overall.
In most cases, bad events are very small (less than 30 data points), but there are exceptions (such as the event on the bottom right of Figure~\ref{fig:sim_examples}).
Bad events tend to have
a highly inaccurate $\vu{z}$, but of course
there is no way to know that without the ground truth.
This highlights the limitation of single-spacecraft measurements: Even when measuring a perfectly time-static and 2D plasma, the spacecraft cannot determine the true $\vu{z}$ and $A$. It can only test whether a given interval satisfies the hypothesis of being 2D and having an $A$ with a single stationary point given a certain $\vu{z}$ by checking if certain necessary conditions are approximately met.
Without a measurement of the magnetic field gradient,
there are no sufficient conditions.
Additional context may be gained by comparing the event's measurements to the surrounding measurements,
which is a potential area of future research to improve single-spacecraft SMFR detection.

\section{Application to \emph{Wind} Data} \label{sec:applicationtowind}

We applied our new algorithm, described in Section~\ref{sec:implementation}, to 27 years of \emph{Wind} data \citep{wilson_quarter_2021} from years 1996-2022.
Each year was processed separately. For magnetic field measurements, we used the Magnetic Fields Instrument (MFI; \citet{lepping_wind_1995}) 3-second vector magnetic field measurements. For measurements of the bulk plasma parameters, we used the proton moments from the 3DP instrument \citep{lin_three-dimensional_1995} computed on-board at a 3-second cadence.
To calculate gas pressure $p$ and {\alf} speed $v_A$, we only included the proton contribution
due to the lack of continuously available quality electron and alpha particle moments.
We used linear interpolation to bring the two datasets onto a set of consistently spaced points,
allowing for a maximum of a 6-second gap between the points used for interpolation.
Points where missing values existed were recorded and then the remaining gaps were filled using linear interpolation.
A number density less than 1 per cc was considered missing, since the quality of the plasma measurements is poor when the measured number density is very low.

The 3DP measurements appear to occasionally have massive spikes, so before interpolating each property onto the consistently spaced points,
we applied a spike removal algorithm based on \citet{roberts_algorithm_1993}. Each point in a 100-point window is marked bad if it is distanced from the mean
of the window by more than six standard deviations.
We repeat this twice rather than iteratively processing each window
as done by \citet{roberts_algorithm_1993} to simplify the algorithm
and make it possible to compute in parallel. Additionally, due to limited telemetry, there is some digitization effect
present in the 3DP data which introduces sudden jumps to the data. To alleviate that, we applied a 5-point running average to the plasma parameters.

After preparing the dataset, we applied the detection algorithm with 195 logarithmically spaced sliding windows
from $10$ (30 seconds) to $10^5$ data points (approximately 3.5 days). The largest window sizes are just for thoroughness: the important range is up to $10^5$ seconds (order of days), since that is the typical order of magnitude of CME durations at 1 au.
We do not expect to see SMFRs larger than CMEs.
A total of 594,857 events were detected, but after restricting the list to only those with fewer than 10\% of the interval containing missing values, 512,152 remained.

Despite the large volume of high-resolution data, our implementation was able to process each year of data in only about a few minutes
on a consumer-level GPU (Nvidia GeForce RTX 3090). Most of the time is spent on the GS reconstruction, without which a year of data can be processed in less than a minute. The high-performance implementation with GPU acceleration made it possible to do much more than could be easily done with the original implementation.
We did not use any supercomputer resources and the improved algorithm also provides GS reconstructions.

\begin{figure}
    \centering
    \includegraphics[width=.8\textwidth]{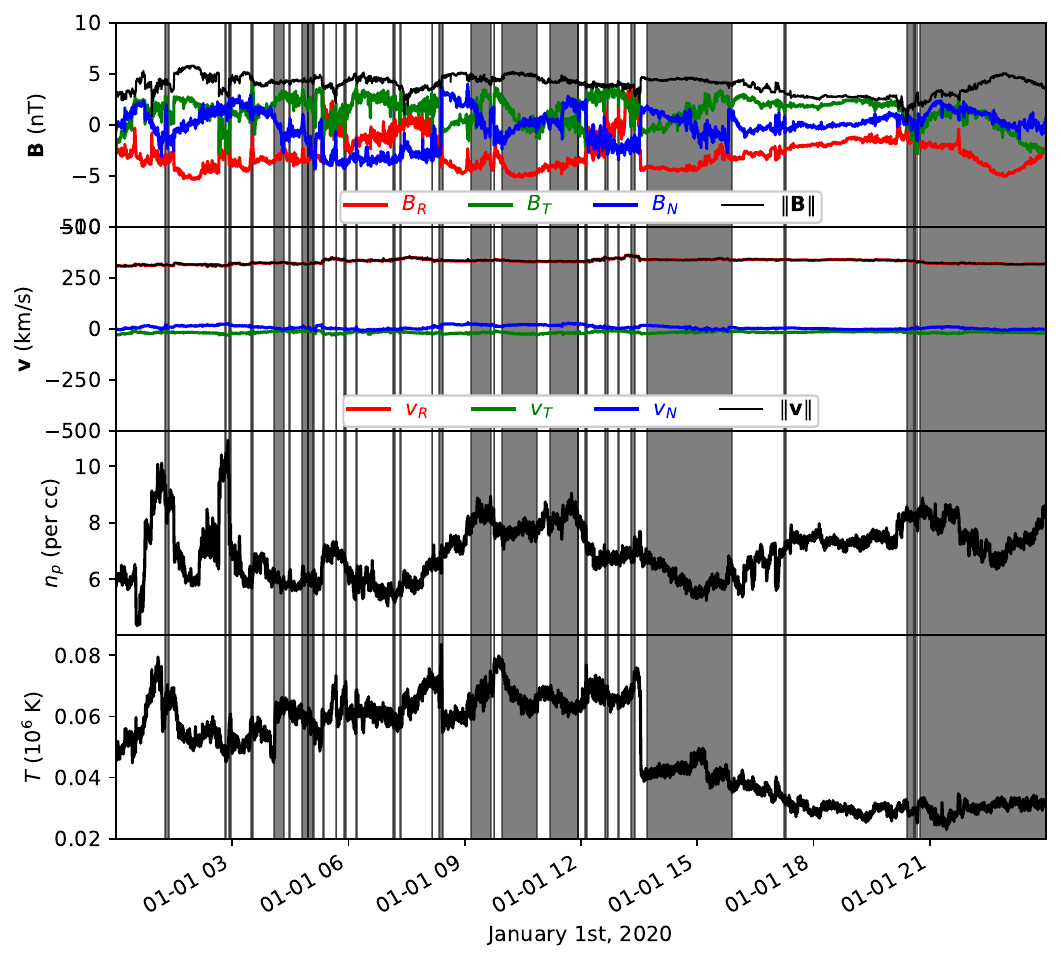}
    \caption{Example of input measurements for a single day, with event intervals shaded gray.}
    \label{fig:example_timeseries}
\end{figure}

Figure~\ref{fig:example_timeseries} displays an example
of the data used for detection for a single day.
This figure is representative of typical quiet solar wind conditions.
Although on average, the magnetic field is aligned with
the Parker spiral, there are usually significant rotations
in the magnetic field direction.
Nearly half of the total
time is detected as a flux rope.
Unlike ICMEs, most of the flux ropes do not have
extremely high magnetic field strength, nor do they necessarily
have decreased proton temperature,
nor do they appear to have any expansion signature (as the velocity does not change much). The sizes of the SMFRs
vary significantly, ranging from less than a minute
to more than an hour. The larger ones are less frequent but occupy a significant portion of the total time; the smaller ones occur in larger numbers but don't compose the majority
of the solar wind.

\begin{figure}
    \centering
    \includegraphics[width=.49\textwidth]{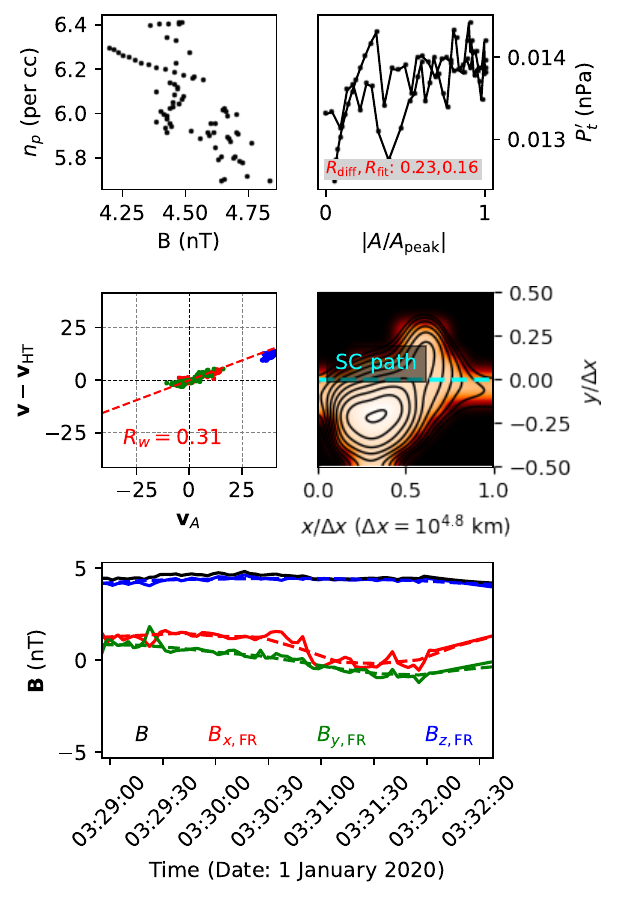}
    \includegraphics[width=.49\textwidth]{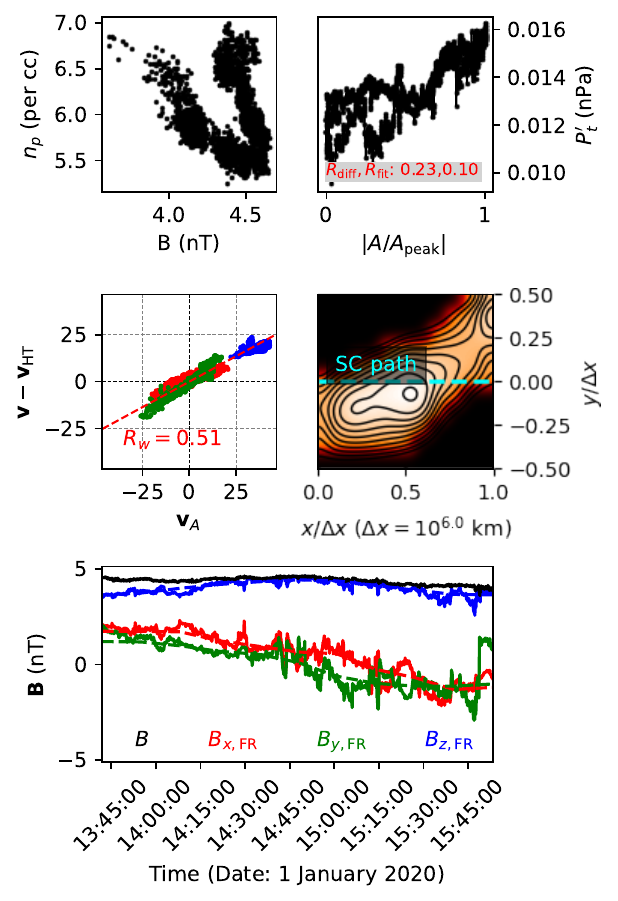}
    \caption{Two examples of real events detecting from \emph{Wind} data. The first event (left) is a relatively small event while the second event (right) is a relative large event.
    Top left: Relation between $B$ and $n_p$.
    Top right: $P_t'$ as a function of $A$ (Appendix~\ref{sec:theory}).
    Center left: Relationship between {\alf} velocity (as a vector) and the velocity fluctuations.
    Center right: Reconstructed cross section with  contours at equally spaced values of the flux function $A$. 
    Bottom: Measured magnetic field components and strength (solid lines) and model magnetic field components (dashed lines) in the flux rope coordinate system.}
    \label{fig:real_examples}
\end{figure}

Figure~\ref{fig:real_examples} shows two examples outputs from our detected algorithm (one small, one large).
Note that both events have $|R_w| > 0.3$, but are inconsistent with an interpretation of them as {\alf} waves due to the nonzero change in $B_z$
and the anticorrelation between $B$ and $n_p$ \citep{vellante_analysis_1987}.
The effect in the field aligned flow can be seen in the green points in the velocity scatter plot, which change sign along with $B_y$.
Even though $R_\mathrm{diff}$ and $R_\mathrm{fit}$
are high compared to the thresholds used by previous studies, the model fits the data very well. The reconstructed cross section shows that GS reconstruction applied to the measurements
reveals closed transverse field lines,
validating the flux rope nature of the events.
Unlike CMEs, neither the small event (about 3 minutes long) nor the long event (about 2 hours long) have an elevated magnetic field strength.
However, many of the detected events do have large changes in the field strength, although most do not exhibit other signatures of CMEs such as velocity expansion signature or very low temperatures.
It seems that the change of magnetic field strength depends on the geometry of the flux rope.
Unlike CMEs, SMFRs do not need large changes in magnetic field strength because they are usually not force-free.

In the following sections, we statistically analyze various aspects of the new database of events.
For additional context, we used the classification scheme introduced by \citet{xu_new_2015}
to distinguish SMFRs in ejecta, sector reversal, streamer belt origin, and coronal hole origin plasma streams.
Since the plasma quantities in \citet{xu_new_2015} are calibrated for 1 hour averaged OMNI2 data, which is primarily based on
\emph{Wind}'s Faraday cup instrument SWE,
we evaluated the solar wind classification based on SWE data downsampled to 1 hour,
then evaluated each event's classification as the the nearest hourly classification.

\section{SMFR Size and Occurrence} \label{sec:spatialtemporal}

\subsection{Size Distribution}

\begin{figure}
    \centering
    \includegraphics[width=.49\textwidth]{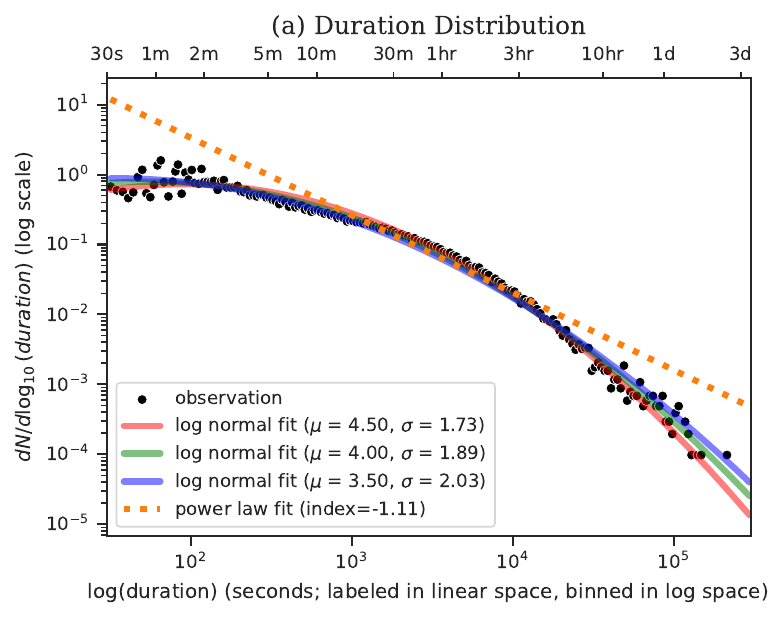}
    \includegraphics[width=.49\textwidth]{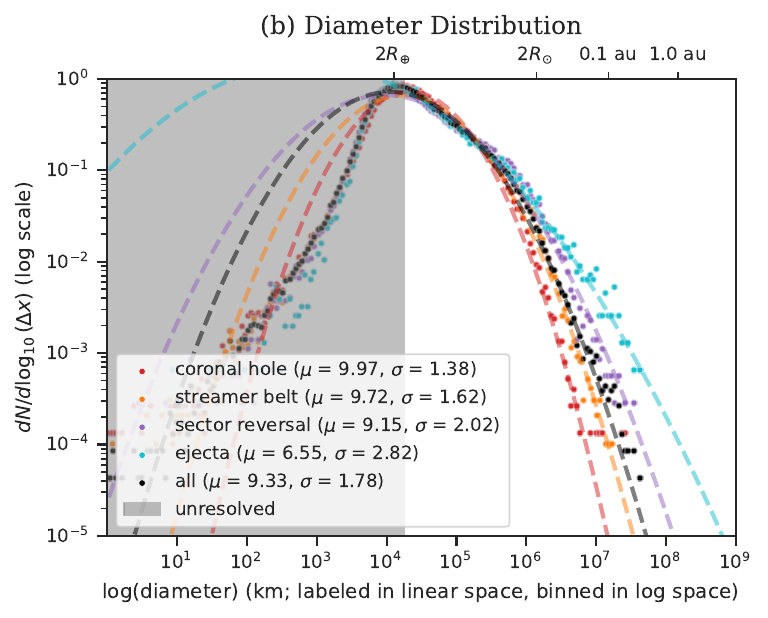}
    \caption{(a) Histogram of the duration (represented as a scatter plot to comparing with the fitted curves). Each black point in the scatter plot represents the number of events detected for a particular sliding window scaled by the bin separation to make it represent the estimated probability density. The points in the scatter plot are centered at the geometric mean of the corresponding window size and the window size after it so that it is centered between the two values in log space. The three curves are log-normal distribution fits with different values of $\mu$. Note that $\mu$ and $\sigma$ are location and scale parameters for the log-normal distribution, not the mean and standard deviation. They are defined so that the natural logarithm of a log-normal distribution is normally distributed with mean $\mu$ and standard deviation $\sigma$. The pink line is a power law fit to the range between 10 minutes and 6 hours.
    (b) Diameter distribution following the same format as (a).
    The shaded region covers diameters that are below $(\SI{600}{\kilo\meter\per\second})(\SI{30}{\second})$. Flux ropes with diameters below this cutoff and orientations perpendicular to the radial direction can be too short for the smallest sliding window.
    The distribution for each solar wind type is also plotted separately with its own fit.}
    \label{fig:size_distr}
\end{figure}

Figure~\ref{fig:size_distr} (a) displays the distribution of the
duration of the detected events. Over the range from 10 minutes to 6 hours, it appears to exhibit a power law similar to the result in the original catalog. However, with the added orders of magnitude, a deviation from a simple power law becomes apparent, and there is significant curvature.
Looking closely at the power law portion of the distribution from our figure as well as the figures in in \citep{hu_automated_2018}, there is already a slight curvature even for the limited range. Our expanded range of durations makes the deviation of the power law obvious.
Log-normal distributions are very common in nature and in the solar wind. It is very common for a log-normal distribution to be mistaken for a power law distribution when viewed over a limited range. This motivates fitting the data to a log-normal distribution, which we demonstrate in Figure~\ref{fig:size_distr} (a). Clearly, the log-normal distribution provides a superior fit. However, the peak of the distribution is below 30s,
the smallest sliding window used in the detection process.
Therefore, the distribution is not fully resolved.
Moreover, a fit of similar quality is attained for an arbitrary choice of the peak as shown in Figure~\ref{fig:size_distr} (a).
Therefore, we cannot determine the exact parameters of the log-normal distribution.
Still, a log-normal distribution clearly fits the data very well
over the resolved range.

The only significant deviation from log-normal in Figure~\ref{fig:size_distr} (a) appears to be a discontinuity around 1 minute. Considering that 1 minute and 30 seconds corresponds to 30 data points,
this may be a consequence of the unreliability of SMFRs detected from such a small number of data points.
The distribution appears to continue to increase all the way down to the smallest scales available from \emph{Wind} plasma measurements. 
Plasma data of higher cadence and quality is thus essential to study SMFRs
at even smaller scales.

In Figure~\ref{fig:size_distr} (b), we consider the spatial scale distribution.
We estimated the diameter of the flux rope as the circle with area equivalent to the closed region in the reconstructed cross section. The spatial width $\Delta x$ of the cross section was estimated using the derived orientation $\vu{z}$ (to determine $\vu{x}$, the direction of motion through the cross section; Figure~\ref{fig:coords}), velocity $\vFR$, and duration $\Delta t$ as $\Delta x = \abs{\vFR \vdot \vu{x}} \Delta t$. This accounts for the lengthening of the observed duration resulting from the angle between the orientation and 
 velocity. If it is perpendicular, then this just reduces to $\|\vFR\| \Delta t$. Once $\Delta x$ is known, the pixels in the reconstructed cross section each have area $(\Delta x/11)^2$, so the area can be estimated by just adding the areas of the individual pixels contained in the closed region of the cross section.
The diameter can be estimated using $\mathrm{Area} = \pi r^2 = \pi d^2/4 \implies d = \sqrt{4\mathrm{Area}/\pi}$.

Figure~\ref{fig:size_distr} plots the diameter distribution separately for each solar wind type. In all cases, a log normal distribution appears to fit the data well.
Besides ejecta, all of the solar wind types have similar log normal distributions, except that slower solar wind types
are more likely to have larger SMFRs than faster ones.
This is consistent with previous studies that have found that
flux tubes/ropes are larger in the slow solar wind
\citep{borovsky_flux_2008,hu_automated_2018}.

The diameter distribution in Figure~\ref{fig:size_distr} (b)
does not have a hard cutoff as in Figure~\ref{fig:size_distr} (a),
but instead has a smooth cutoff.
The reason SMFRs below the cutoff imposed by the temporal
scale limitation can sometimes be detected
is that the duration depends not only on the diameter, but also the impact parameter and the angle between $\vu{z}$ and $\vFR$. For a given diameter and impact parameter, if $\vu{z}$ is not perpendicular to $\vFR$, the duration will be longer.
The highest typical velocity is approximately \SI{600}{\kilo\meter\per\second}, so for a minimum window size of \SI{30}{\second}, one would expect the distribution to be inaccurate below $(\SI{600}{\kilo\meter\per\second})(\SI{30}{\second})$ (the shaded region in Figure~\ref{fig:size_distr} (b)), which is almost precisely where the distribution begins to decrease.
Therefore, it is likely that even smaller scales exist. We cannot determine the peak of the distribution.

A small but nonzero fraction of the events have diameters below $\SI{100}{\kilo\meter}$, which is approximately the transition from MHD scales to kinetic scales. However, these events are not reliable. Besides the inapplicability of the MHD approximation at such small scales, the temporal scale limitation means that such small events can only be detected when $\vu{z}$ at a very small angle from $\vFR$. For $\|\vFR\| = \SI{400}{\kilo\meter\per\second}$,
and sliding window size \SI{30}{\second},
we must have that $\SI{400}{\kilo\meter\per\second} \SI{30}{\second} \sin(\theta) = \SI{100}{\kilo\meter}$
where $\theta$ is the angle between $\vu{z}$ and $\vFR$. This yields $\theta \approx \SI{0.5}{\degree}$, which is far beyond the precision afforded by single-spacecraft measurements. Thus, we cannot confirm whether flux ropes exist in the solar wind below MHD scales.

\subsection{Lack of Variation over Solar Cycle}

\begin{figure}
    \centering
    \includegraphics[width=0.9\textwidth]{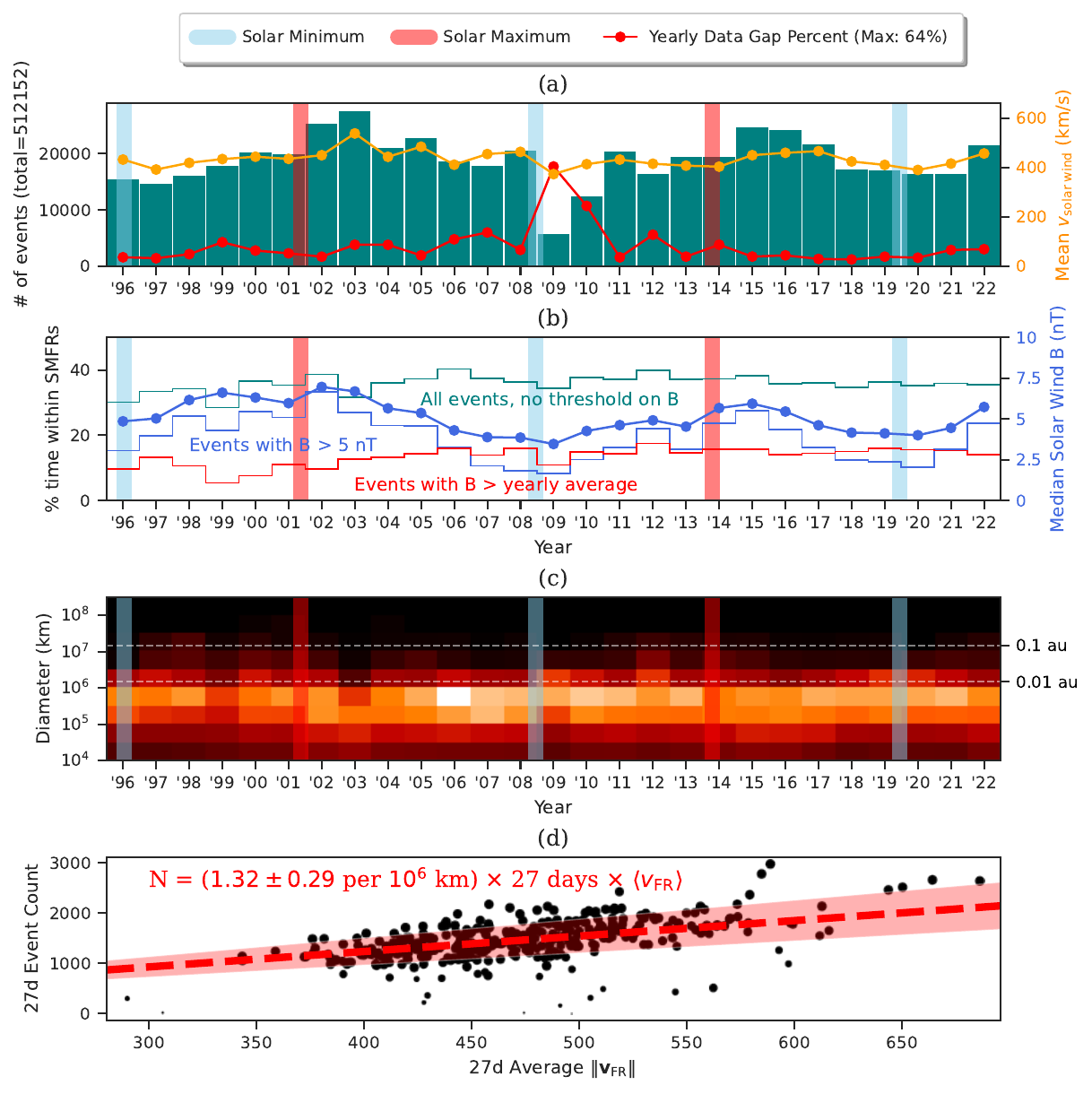}
    \caption{(a) Displays the number of events in each year displayed as a bar plot. Overplotted in (a) are the relative variations of the yearly averaged solar wind velocity as measured by SWE (including the entire year, not just SMFR intervals).
    Additionally, the solar maxima and minima are labeled.
    This plot demonstrates that without a threshold fixed to a particular value, the variations in SMFR count correspond to velocity variations or data gaps, not solar activity.
    (b) Filling factor in each year containing SMFRs (corrected for data gaps). Also included is the yearly average magnetic field strength and the filling factor with two alternative $B$ thresholds.
    (c) Filling factor in each year containing SMFRs within particular ranges
    of diameters. Brighter bins indicate a higher percentage of the year contained
    within SMFRs of diameters within the bin's range of diameters.
    (d) Scatter plot including linear regression between each 27 day period's average $\|\vFR\|$ and number of events. The dashed line is the fitted regression and the shaded region contains the 4$\sigma$ uncertainty. Smaller points have more data gaps. The periods with no SMFRs are not included in the scatter plot.}
    \label{fig:yearly_variation}
\end{figure}

Figure~\ref{fig:yearly_variation} (a) shows the number of events
detected from each year of data.
The number is not constant, but it doesn't exactly correspond to solar activity, either.
The points in time corresponding to solar maxima and minima are displayed as red
and blue vertical bars. The peak SMFR number is not exactly at solar maximum
and the minimum SMFR number is not exactly at solar minimum.
This implies that the SMFR counts are not directly related to solar activity,
but they are related to the solar cycle.
They appear to peak in the declining phase of the solar cycle.
In fact, it is well known that the solar wind speed peaks
during the declining phase of the solar cycle.
This motivates the inclusion of the variation of the yearly average solar wind speed
in Figure~\ref{fig:yearly_variation} (a),
showing that the number of SMFRs is directly proportional to the average speed.
The only significant deviations are during years that have major data gaps (also included in Figure~\ref{fig:yearly_variation} (a)).
Figure~\ref{fig:yearly_variation} (d) confirms that the same trend applies on a timescale of 27 days (about a synodic solar rotation).
This implies that the filling factor of the SMFRs is nearly constant over time. If the filling factor is constant and the velocity doubles, the number of SMFRs should double. Mathematically:
\[\dv{N}{t} = \dv{N}{x} \dv{x}{t} = \dv{N}{x} v \approx \mathrm{constant} \times v \implies \dv{N}{x} \approx \mathrm{constant}\]
A quantitative estimate is provided by fitting the number of events to the average velocity in a given 27 day period in Figure~\ref{fig:yearly_variation} (d), yielding an estimate of ${\sim} 1$ SMFR per $10^6$ km.

If the radial density is constant, then the filling factor containing SMFRs should also be constant. Figure~\ref{fig:yearly_variation} (b) shows the yearly integrated duration of all of the SMFRs in a given year divided by one year. In other words, it shows the percentage of a year contained within the events. From this figure, it is apparent that the temporal variation is minimal. and clearly has no correlation to the sunspot number. The variation imposed by the change of yearly average bulk solar wind speed in Figure~\ref{fig:yearly_variation} (a) is also eliminated in Figure~\ref{fig:yearly_variation} (b).
 It appears that approximately 35\% of the solar wind contains SMFRs for the entire solar cycle. This may even be an underestimate, due to the fact that the smallest scales are not resolved (Section~\ref{sec:spatialtemporal}) and because some of the assumptions required in the detection process may not apply to all SMFRs. In contrast, applying the fixed 5 nT threshold results in a strong solar cycle dependence, with the filling factor dropping severely when the yearly average B drops below 5. Yet using the yearly average B instead of 5 nT gives virtually the same filling factor trend as using no threshold at all, except that the value is halved. This shows that the solar cycle dependence
 for the majority population is artificially imposed by the fixed 5 nT threshold.

Figure~\ref{fig:yearly_variation} (c) shows the dependence on size
by plotting the filling factor (percentage of time) contained in SMFRs within a given year and range of sizes. From here it becomes apparent that despite the probability density of smaller SMFRs increasing below the smallest resolved size,
the filling factor occupied by these extremely small SMFRs is very low. Most of
the time is occupied by SMFRs of size between $10^5$ and $10^6$ km.
These do not exhibit any sort of clear solar activity dependence. In fact there is a weak anticorrelation with solar activity, where the filling factor occupied by SMFRs around $10^6$ km seems to peak at solar minimum. However, looking closely at the top portion of Figure~\ref{fig:yearly_variation} (c),
events of diameter above ${\sim}\SI{0.01}{au}$ appear to be significantly more common during 
solar maximum and have a strong correlation with solar activity.
Considering CMEs tend to be on the order of 0.1 au,
these would still be considered SMFRs.
Due to the solar activity dependence and large size,
this subpopulation is much more likely to have a solar origin.

\subsection{Do SMFRs Cluster Around the Heliospheric Current Sheet?}

\begin{figure}
    \centering
    \includegraphics[width=.5\textwidth]{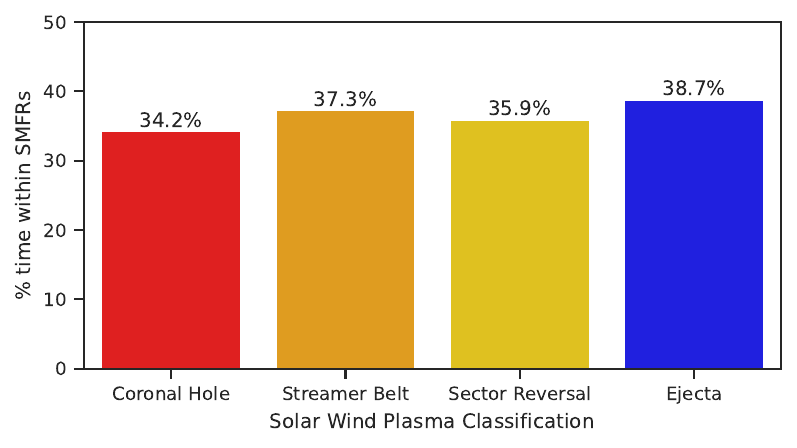}
    \caption{Filling factor for each category within SMFRs (corrected for data gaps).}
    \label{fig:occurrence_by_origin}
\end{figure}

Figure~\ref{fig:occurrence_by_origin} shows the filling factor for each solar wind classification (based on \citet{xu_new_2015}). In both of the slow solar wind categories, about 43\% of the time is filled with SMFRs. The slight drop in percentage for coronal hole origin plasma might be a result of the higher speed resulting in less events fitting in the sliding windows compounded with the known fact that SMFRs tend to be slightly smaller in faster solar wind \citep{borovsky_flux_2008,hu_automated_2018}. 
That the filling factor is independent
of solar wind type is very surprising, because \citet{cartwright_heliospheric_2010} found that SMFRs cluster around the heliospheric current sheet (HCS), and this conclusion was in agreement with the analysis of \citet{hu_automated_2018}.
If that were the case, we should have seen the highest density of SMFRs at sector reversal regions.
In this subsection, we demonstrate that the apparent tendency of SMFRs to cluster around the HCS is artificial.

In their Section 7, \citet{hu_automated_2018} analyzed the distribution of the number of days between
the detected SMFRs and the nearest HCS crossing
in order to compare to a similar analysis by \citet{cartwright_heliospheric_2010}. It was found that the distribution is peaked at 1 day after the nearest HCS crossing
and that the SMFRs tend to cluster around the HCS boundaries.
However, we point out here
that looking solely at the distribution of days between SMFR times and the nearest HCS crossing is prone to statistical bias. Most of the measurements are close to the HCS, so the distribution for SMFRs must be compared to the distribution of the measurements. The distribution of 
 the SMFR distance to the HCS is only meaningful by itself if the distance of the measurements from the HCS
are uniformly distributed.

\begin{figure}
    \centering
    \includegraphics[scale=0.6]{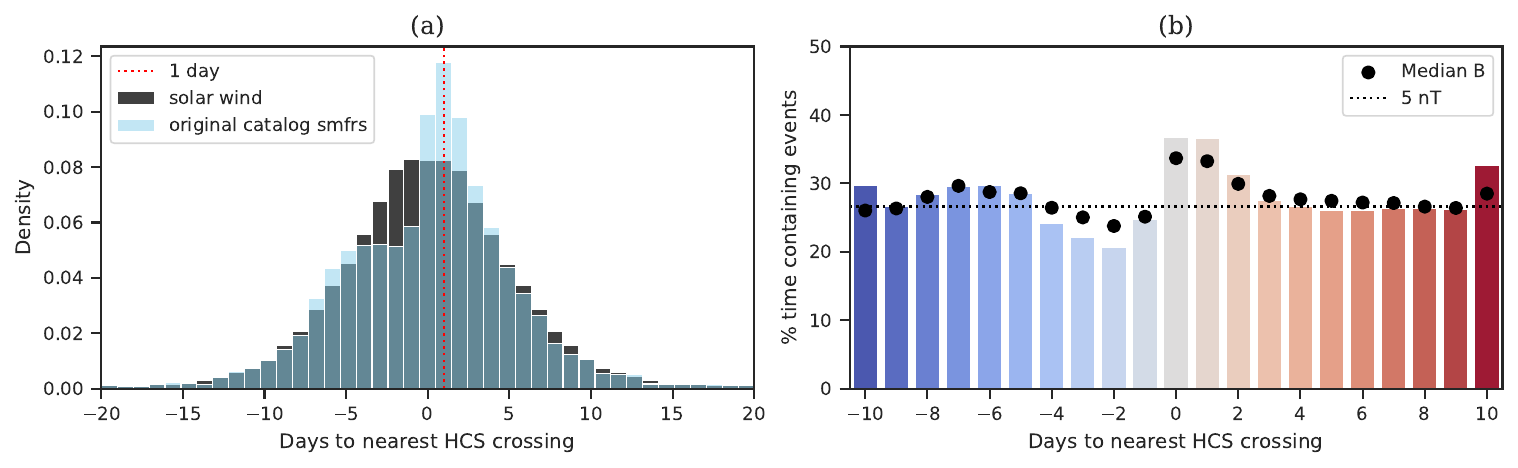}
    \\[\smallskipamount]
    \includegraphics[scale=0.6]{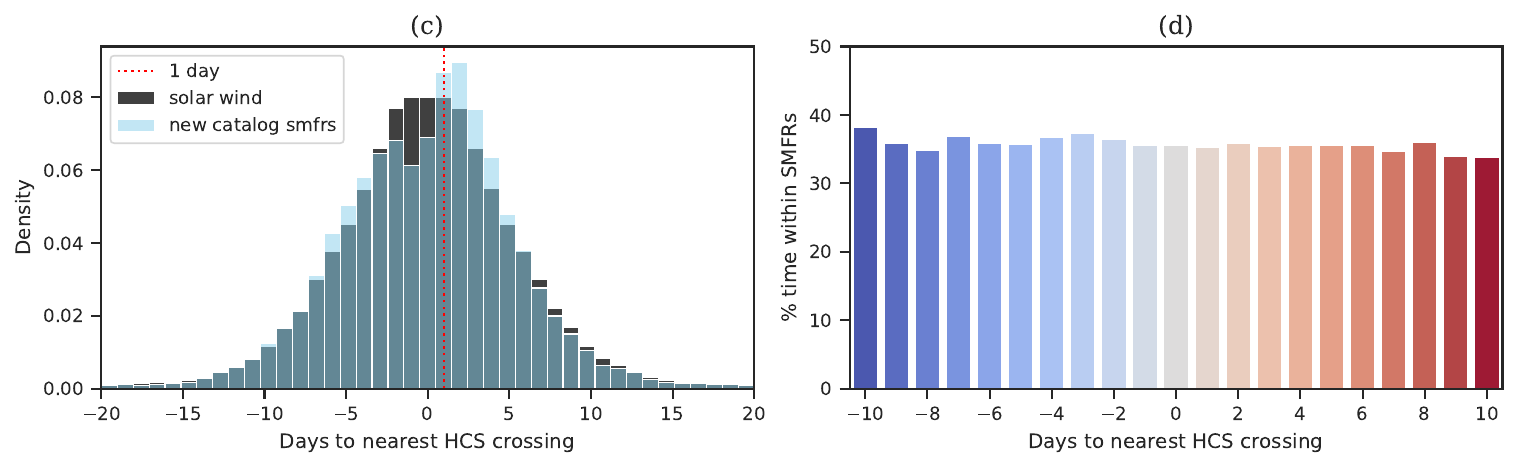}
    \caption{(a) Histogram of distance from nearest HCS crossing for original catalog's events and for uniformly spaced timestamps. The black bars represent the distribution of distance
from the HCS to points evenly spaced in time for the time interval of the original catalog.
The blue bars represent the distribution of the distance from the HCS for the SMFRs in their catalog. The dotted red line demonstrates that the SMFR distribution peaks at 1 day from the HCS. (c) and (d): Same as (a) and (b) but using the data for the new list of events. Additionally, (d) is corrected for data gaps but not (b) because the data gaps are less significant in the original catalog dataset and because it used a different instrument.}
    \label{fig:hcs_distr}
\end{figure}

In order to determine whether SMFRs are more common closer to the HCS crossings,
we calculated the distances to the nearest HCS crossing for the measurements used for detection in addition
to the distances for the detected SMFRs. 
Like \citet{hu_automated_2018}, we use L. Svalgaard's list of HCS crossings\footnote{https://svalgaard.leif.org/research/sblist.txt}.
In Figure~\ref{fig:hcs_distr} (a), we compare their HCS distance distribution to the SMFR HCS distance distribution.
From this figure, it appears
that the filling factor of SMFRs
is constant far from HCS crossings,
reduced shortly before,
and elevated shortly after.
However, the changes are strongly correlated with changes in the average $B$ as a function of distance from the HCS crossing as shown in the figure, suggesting the possibility of further statistical bias.

In Figure~\ref{fig:hcs_distr} (b),
we plotted the total duration of the SMFRs that are a given number of days from the nearest HCS crossing by
the total time that \emph{Wind} is that distance from the nearest HCS crossing (filling factor).
For most of the distances, it is essentially a constant 25\% with minimal variation, so there is not a strong tendency to be close to the HCS crossings. However, there is a slight decrease to around 20\% right before the HCS crossing, followed by an increase to around 35\% after the HCS crossing.
As observed by \citet{hu_automated_2018}, the SMFRs are more likely to occur approximately one day after the HCS than during or before the HCS crossings. Figure~\ref{fig:hcs_distr} (b) also confirms the dip in the proportion of SMFRs right before an HCS crossing. The break point from a uniform distribution is within 6 days, which is close to the typical distance between HCS crossings during quiet times (considering a 4-sector HCS over the 27 day solar rotation).

\begin{figure}
    \centering
    \includegraphics[width=.8\textwidth]{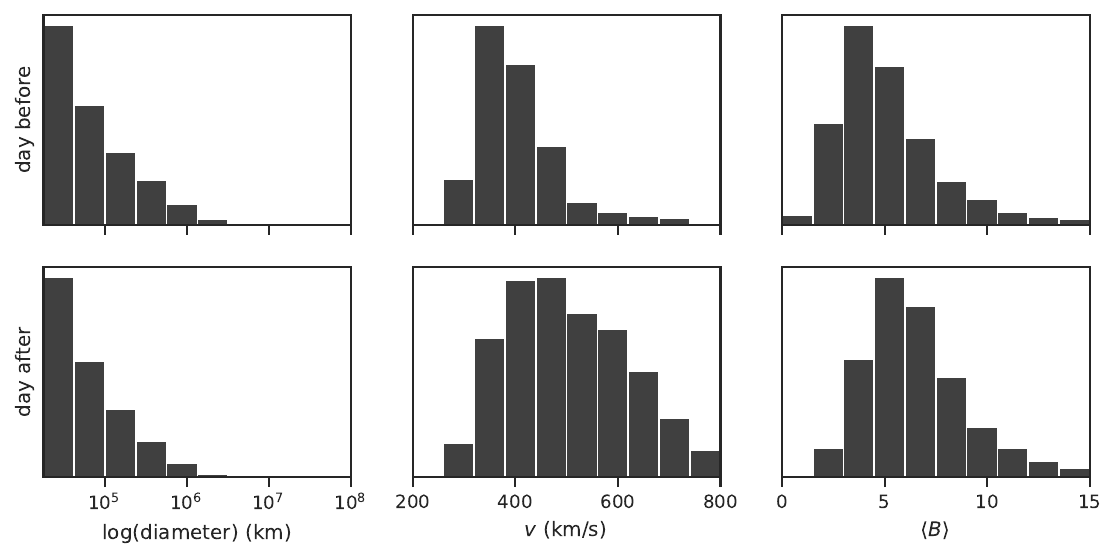}
    \caption{Comparison SMFR diameter, speed ($\|\vFR\|$), and magnetic field strength $B$ distributions
    before and after HCS crossings. The diameter histograms is only shown for the fully resolved range, and the other histograms are limited to typical ranges. The y axis is the proportion of samples in each  bin with the y limit set to 50\%.
    The y labels are not shown because the focus is the difference in shape of the distribution between one day before and one day after the nearest HCS crossing. Note in particular that the diameter histograms are approximately the same whereas the velocity histogram for one day after has a significantly higher proportion for higher velocities than one day before, and that the magnetic field strength has a significantly higher proportion above $\SI{5}{\nano\tesla}$ a day after than a day before.}
    \label{fig:hcs_crossing_properties}
\end{figure}

Figure~\ref{fig:hcs_distr} (c) is of the same format as Figure~\ref{fig:hcs_distr} (a), but using the new catalog.
It is qualitatively the same, although quantitatively, the
deviations of the SMFR distribution from the measurement distribution are less pronounced.
However, the filling factor based on the new catalog (Figure~\ref{fig:hcs_distr} (d)) exhibits a different result from the original catalog: the filling factor is essentially the same for all distances!
The disagreement with the original catalog can be explained by the magnetic field strength distributions in Figure~\ref{fig:hcs_crossing_properties}.
Before the HCS crossing, the average magnetic field strength is below $\SI{5}{\nano\tesla}$. After the HCS crossing, it is above $\SI{5}{\nano\tesla}$. Due to the fixed threshold of $\SI{5}{\nano\tesla}$,
this resulted in more events being detected after HCS crossings and less before HCS crossings in the original catalog.

If the filling factor of SMFRs does not depend on distance from HCS, why is there a difference in the number of SMFRs before and after the HCS crossing?
The answer lies in Figure~\ref{fig:hcs_crossing_properties}, which shows that while the size distribution of SMFRs does not differ much before or after HCS crossings, the velocity and magnetic field strength distributions are higher after the HCS crossing.
This is because corotating interaction regions (CIRs),
interfaces between the fast and slow solar wind,
are known to ``catch up'' to the HCS
very often
\citep{borrini_solar_1981,balogh_cir_1999,huang_coincidence_2016,potapov_current_2018,liou_characteristics_2021}.
Faster solar wind streams
have higher magnetic field strength and velocity.
A tendency for the number of SMFRs to increase a day after
HCS crossings due to the increase in velocity.
For example, if the velocity were two times higher, we would expect double the number of SMFRs if the same filling factor is the same.
The small increase in filling factor in Figure~\ref{fig:hcs_distr} (d) right before the HCS crossing may be due to the fixed minimum sliding window length: A slightly higher proportion of the total duration will be filled with events when the velocity is lower, since more events can fill the sliding windows due to having a longer duration.

In summary, the radial density and filling factor of SMFRs is independent of distance from the HCS.
All of the observations indicating otherwise
can be understood as follows:
The velocity would be expected to increase after crossing the HCS and passing the CIR, usually after less than a day \citep{liou_characteristics_2021}. This explains the peak of the distribution a day after the HCS crossing rather than zero days, as pointed out by \citet{hu_automated_2018} and confirmed in the above analysis.
The higher velocity after the HCS results in more SMFRs being detected without affecting the filling factor.
The reason the events in the original catalog have a difference in filling factor before and after the HCS, not just the number of events, is because of the $\avg{B} > \SI{5}{\nano\tesla}$ threshold: $B$ increases from below to above the threshold after entering faster solar wind with stronger $B$.
In short, although more SMFRs are observed when the solar wind moves faster, the filling factor of SMFRs is independent of distance from the HCS (as well as solar activity). It is constantly approximately 35\%.
Since the size distribution varies only slightly by distance from HCS or plasma type, this means that radial density is independent of distance from the HCS, too.

\section{Physical Properties} \label{sec:properties}

\subsection{Axial Orientation}

\begin{figure}
    \centering
    \includegraphics[width=.45\textwidth]{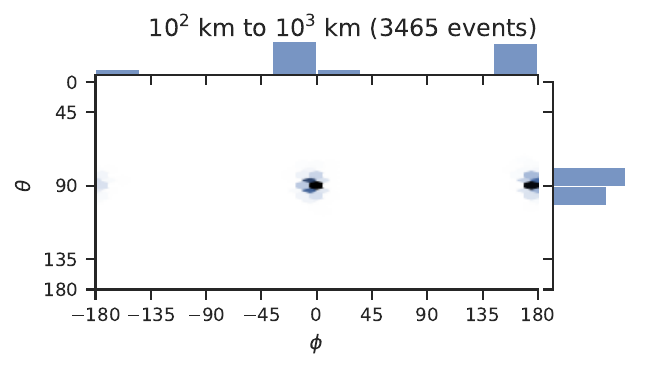}
    \includegraphics[width=.45\textwidth]{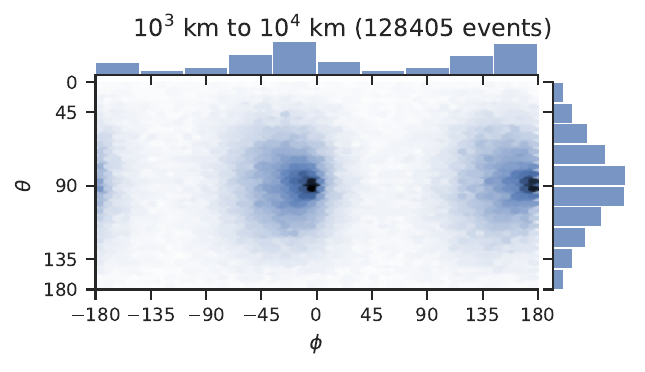}
    \includegraphics[width=.45\textwidth]{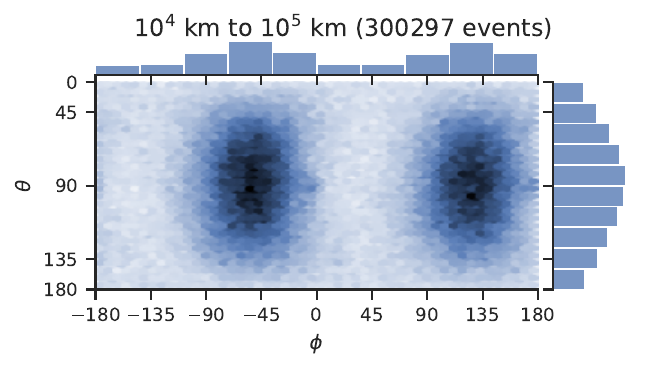}
    \includegraphics[width=.45\textwidth]{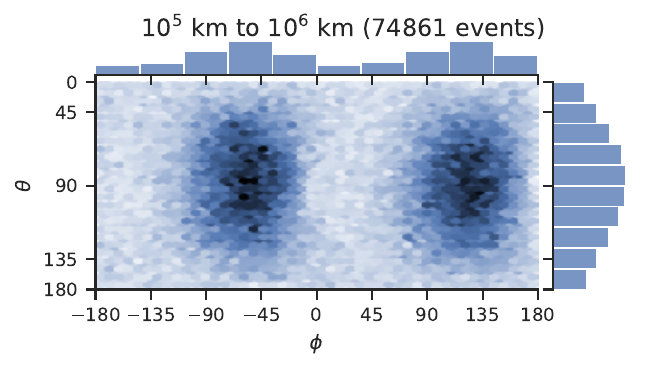}
    \includegraphics[width=.45\textwidth]{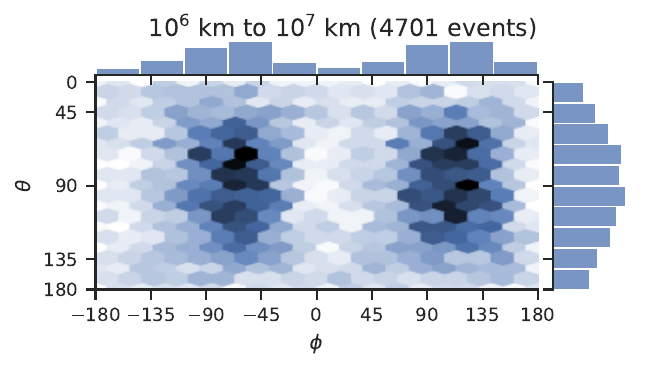}
    \includegraphics[width=.45\textwidth]{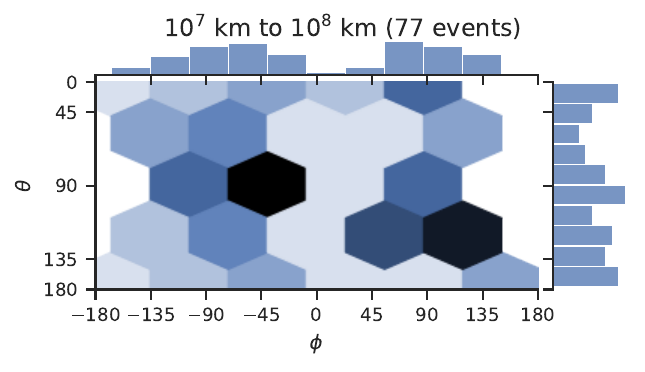}
    \caption{Joint distribution of $\vu{z}$ azimuth $\phi$ (counterclockwise angle from the radial direction away from the Sun) and polar angle $\theta$ (angle from the normal direction in the RTN coordinate system) for various ranges of diameter. In the 2D histograms, darker values mean higher filling factors. The 2D histograms are generated with hexagonal binning to clearly show the shape. Furthermore, the vertical binning is by $\cos(\theta)$ instead of $\theta$ (hence the uneven spacing) because randomly oriented unit vectors naturally tend to values of $\theta$ closer to $\SI{90}{\degree}$ (by a factor of $\sin(\theta)$) but have uniformly distributed $\cos(\theta)$. The sides show the histograms of $\phi$ (on top) and $\theta$ (on the right).}
    \label{fig:orientation_distr}
\end{figure}

Figure~\ref{fig:orientation_distr} illustrates the distribution of the axial orientation in the radial-tangential-normal (RTN) coordinate system converted to spherical coordinates. In terms of the components of $\vu{z}$ in RTN coordinates, $\phi \equiv \mathrm{arctan2}(z_T, z_R)$ is the azimuth angle and $\theta \equiv \arccos(z_N)$ is the polar angle. $\phi = 0$ means radially outward from the Sun, and $\theta = 0$ means in the normal direction
(which is approximately northward), while $\theta = \SI{90}{\degree}$ means in the RT plane.
Parker spiral alignment would have $\theta = \SI{90}{\degree}$ and $\phi \approx -\SI{45}{\degree}$ when the IMF has positive polarity (away from the Sun) and $\phi \approx \SI{135}{\degree}$ when it has negative polarity (towards the Sun). In Figure~\ref{fig:orientation_distr}, this appears to be the case for all of the well-resolved scale ranges. Deviations from Parker spiral alignment follow a 2D Gaussian distribution, implying that they are due to random and independent processes (such as errors in determining the orientation or, alternatively, 3D effects such as the tangling of flux tubes into spaghetti as illustrated in \citet{borovsky_flux_2008}).

In Figure~\ref{fig:orientation_distr}, there are peaks in the smallest two ranges
that are shifted about \SI{5}{\degree} clockwise from the (anti)radial direction. Due to the Earth's counterclockwise orbit, the solar wind velocity relative to a spacecraft orbiting along with the Earth has a slight clockwise shift. The sign of the direction depends on the IMF polarity, not the direction of the velocity. Thus, the peaks are velocity aligned orientations. Because the temporal scale distribution continues to increase at our smallest sliding window size of 30s, it is likely that a significant number of SMFRs exist at scales below the lowest well-resolved scale in our event list (approximately $10^4$ km or so). It takes longer for the spacecraft through a flux rope of a given size the smaller the angle between its orientation and velocity. Thus, the many events with diameters below the resolved range would only be detected if their orientation is sufficiently close to the velocity that their duration reaches 30s. Indeed, the peaks virtually disappear when events of diameter lower than $10^4$ km are ignored, which is approximately the cutoff (Section~\ref{sec:spatialtemporal}). This explains the additional peaks close to the velocity direction.

The results in this section demonstrate that the SMFRs at all scales where the orientation distribution can be resolved have a clear tendency to follow the Parker spiral direction. This is in agreement with the results of the original catalog, providing further validation for the extended range of sizes.

\subsection{Field-Aligned Flows in SMFRs} \label{sec:alfvenicity}

\begin{figure}
    \centering
    \includegraphics[width=.6\textwidth]{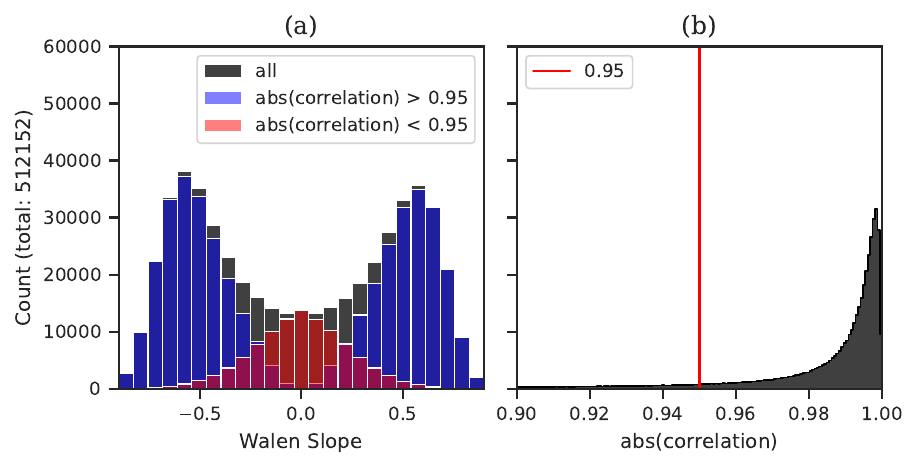}
    \includegraphics[width=.3\textwidth]{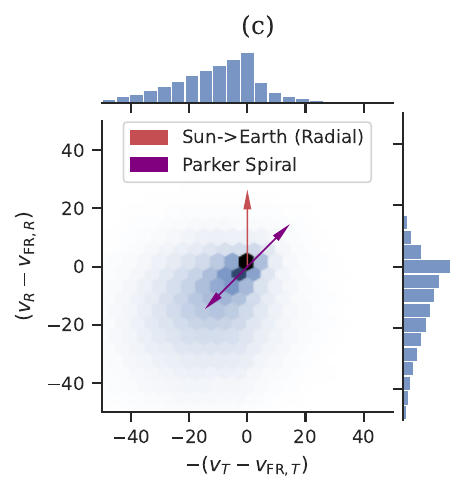}
    \caption{(a) Histogram of {\wal} slope. The black bars include all events, while the blue and red respectively represent events with extremely strong and not so strong correlation between $\vb{v} - \vHT$ and $\vb{v}_A$. A positive (negative) {\wal} slope indicates a flow (anti)parallel to the magnetic field. (b) Histogram of correlation between $\vb{v} - \vFR$ and $\vb{v}_A$. (c) Histogram of velocity difference between $\avg{\vb{v}}$ and $\vFR$. The plot indicates a tendency for the velocity fluctuations to be sunward along the Parker spiral in the flux rope frame of reference.}
    \label{fig:walen_distr}
\end{figure}

The {\wal} slope distribution, measuring {\alf}icity, is plotted in Figure~\ref{fig:walen_distr} (a). In Figure~\ref{fig:walen_distr} (b),
the distribution of the correlation is shown as well.
When the correlation is low, the {\wal} slope tends to be estimated as 0 since a clear linear relationship cannot be found.
For the most part, the results are consistent with the results we obtained by reanalyzing the original catalog (Figure~\ref{fig:original_catalog_walen_distr}).
Most of the SMFRs have non-negligible field-aligned velocity fluctuations that are comparable to but less than the {\alf} velocity.
Both along and opposite to the magnetic field direction,
the absolute value appears to be normally distributed.
An additional population centered around 0 is present but can be explained as a consequence of measurement uncertainties resulting in low correlation for events with weak field-aligned flows resulting in a {\wal} slope tending to zero.
However, unlike the center in Figure~\ref{fig:original_catalog_walen_distr}, the center of the normal distributions in Figure~\ref{fig:walen_distr} (a) appears to be lower than 0.7.
This is most likely because the lack of the 5 nT threshold enables more flux ropes to be detected
that are not in coronal hole origin solar wind streams, hence less {\alf}ic and lower $B$.
This also explains the difference between our result and the result of \citet{borovsky_motion_2020}, since their statistics are based only on {\alf}ic events.

If the velocity fluctuations in our detected events do not average to zero, does that mean that they are actually waves, not propagating structures? Waves can be broadly defined as structures that propagate relative to the so-called background bulk fluid velocity, whatever that is.
There is a difference between the flux rope velocity and the average bulk fluid
velocity within the flux rope.
Figure~\ref{fig:walen_distr} (c) demonstrates that the majority of SMFRs move faster than the average of the velocity within the SMFRs by a finite but sub-{\alf}ic amount, usually less than half the {\alf} speed. Thus from the perspective of the particles within a flux rope structure, most flux ropes propagate away from the Sun along the Parker spiral. In the flux rope frame of reference, the plasma tends to flow sunwards along the Parker spiral. 

The velocity fluctuations, being field aligned,
will have a mean fluctuation that is aligned with the flux rope axis.
Since the velocity fluctuations have a preferred direction, the rest from of the solar wind plasma is not necessarily the average velocity $\avg{\vb{v}}$. Previous studies suggest that the solar wind frame of reference is in fact $\vHT$, the velocity of the advected structures (e.g. \citet{nemecek_what_2020}). The variations in both magnetic field and plasma velocity, as well as variations in properties related to the magnetic structure like density, specific entropy, plasma beta, helium abundance, and electron heat flux are all advected with velocity $\vHT$ \citep{borovsky_motion_2020}.
This suggests that the rest frame is in fact $\vHT$ ($\equiv \vFR$), not $\avg{\vb{v}}$.
With this information in mind, it is more likely that we are dealing with advected structures, not waves.

\begin{figure}
    \centering
    \includegraphics[width=.9\textwidth]{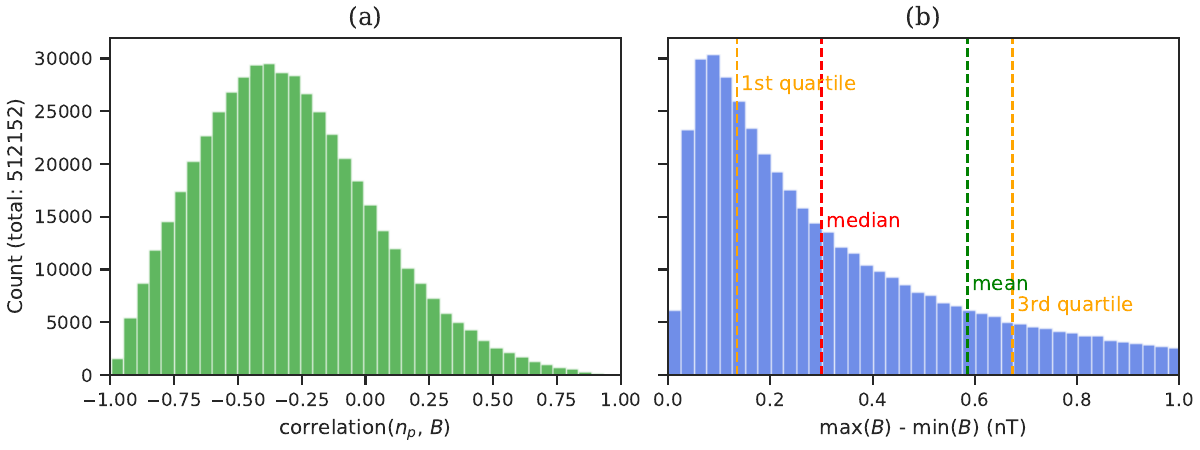}
    \caption{(a) Histogram of correlation between proton number density and magnetic field strength. (b) Histogram of the maximum B minus minimum B in the closed part of the reconstructed cross section. Demonstrates that unlike pure {\alf} waves, SMFRs have anticorrelated $n_p$ and $B$
    as well as a nonzero, measurable change in $B$.}
    \label{fig:NBcorr_B_distr}
\end{figure}

Advected structures are often distinguished from waves by anticorrelated proton density $n_p$ and magnetic field strength $B$, which is not a signature exhibited by {\alf} waves.
\citet{burlaga_microscale_1976} (cited by \citet{cartwright_heliospheric_2010} to justify the exclusion of {\alf}ic fluctuations from SMFR detection) pointed out
`{\alf} waves' from spacecraft observations
have nonzero, measurable fluctuation in $B$,
which cannot be incompressible {\alf} waves but can be compressible {\alf} waves.
\citep{burlaga_magnetic_1970} found
weakly anticorrelation between total thermal pressure
and magnetic pressure on a timescale of ${\sim}$1 hour, evidence for the existence of pressure-balanced
nonpropagating structures.
\citet{denskat_multispacecraft_1977} found evidence that some {\alf}ic
fluctuations may contain tangential discontinuities and other types of static
structures, suggesting they are probably not pure {\alf} waves.
Anticorrelation between $n_p$ and $B$
have been used by previous studies as evidence of the existence of pressure balanced structures to the exclusion of {\alf} waves (e.g. \citet{vellante_analysis_1987,matthaeus_evidence_1990}).
Figure~\ref{fig:NBcorr_B_distr} (a) shows a strong tendency for negative correlation between $n_p$ and $B$. This suggests that the detected SMFRs correspond to the long-observed pressure balanced structures,
not {\alf} waves.
A property expected for incompressible {\alf} waves is a constant magnetic field strength $B$,
but Figure~\ref{fig:NBcorr_B_distr} (b) demonstrates that for the detected events,
the range of magnetic field strength values is usually significantly higher than the measurement uncertainty (which is less than 0.1 nT; not shown here, the histogram of the logarithm shows that the range of magnetic field strength is log normally distributed).
Since $|R_w|$ is usually significantly less than 1,
even pure compressible {\alf} waves cannot
satisfactorily explain the data.
Considering also that the {\alf}ic events have most of the same statistical properties as the non-{\alf}ic events,
it is difficult to consider the {\alf}ic events to be pure {\alf} waves and not flux ropes.

\begin{figure}
    \centering
    \includegraphics[width=.6\textwidth]{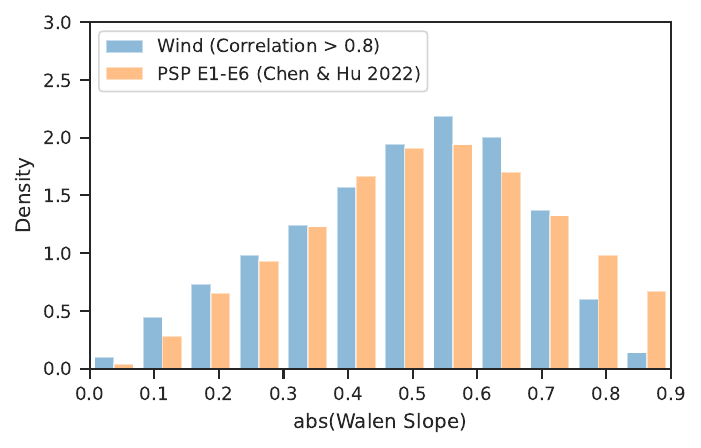}
    \caption{Comparison of absolute value of {\wal} slope with the distribution derived from PSP's first six encounters by \citet{chen_small-scale_2022}. Only the absolute value is shown because most of those PSP encounters were during periods of antiradial magnetic field, so the {\wal} slope from the PSP observations was preferentially positive.
    For the \emph{Wind} data, only samples with at least 0.8 correlation are included in the histogram, so that the criteria is the same as the one used by \citet{chen_small-scale_2022}. Additionally, since we require that the {\wal} slope is less than 0.9, we do not include bins above 0.9. This figure shows that the peak of the distribution is the same. The differences in the other parts of the distribution may be due to the differences in plasma types observed by the two spacecraft.}
    \label{fig:PSPcomparison}
\end{figure}

Considering that the previous understanding is that
SMFRs near the Sun are more {\alf}ic than SMFRs away
from the Sun, it is pertinent to compare our results
to recent findings from PSP.
Figure~\ref{fig:PSPcomparison} compares our derived {\wal} slope distribution
to the one derived by \citep{chen_small-scale_2022} from PSP's first 6 encounters.
The most outstanding feature of this figure is
that the peak is the essentially same at 1 au and at PSP.
However, there are some significant differences away from the peak.
It is unclear whether this difference is due to the radial difference or the difference in plasma types observed by the two spacecraft.
If PSP happened to observe more {\alf}ic solar wind
than non-{\alf}ic solar wind, for example,
then such a difference between PSP and 1 au observations
should occur even without any radial evolution of the {\alf}icity.
Since the peak of the distribution is the same, it seems that there is minimal variation of the {\wal} slope between the inner heliosphere and 1 au, if any. 
Although it is well-known that the ``{\alf}icity'' in the sense of cross helicity decreases away from the Sun, this does not necessarily mean that the ``{\alf}icity'' in terms of {\wal} slope has any radial dependence: the {\alf} speed decreases away from the Sun, which reduces the energy of {\alf}ic fluctuations.
Since cross helicity is related to the energy of the {\alf}ic fluctuations,
this would reduce the cross helicity while leaving the {\wal} slope the same.

\subsection{Magnetic Flux, Twist, and Current Density}

\begin{figure}
    \centering
    \includegraphics[width=\textwidth]{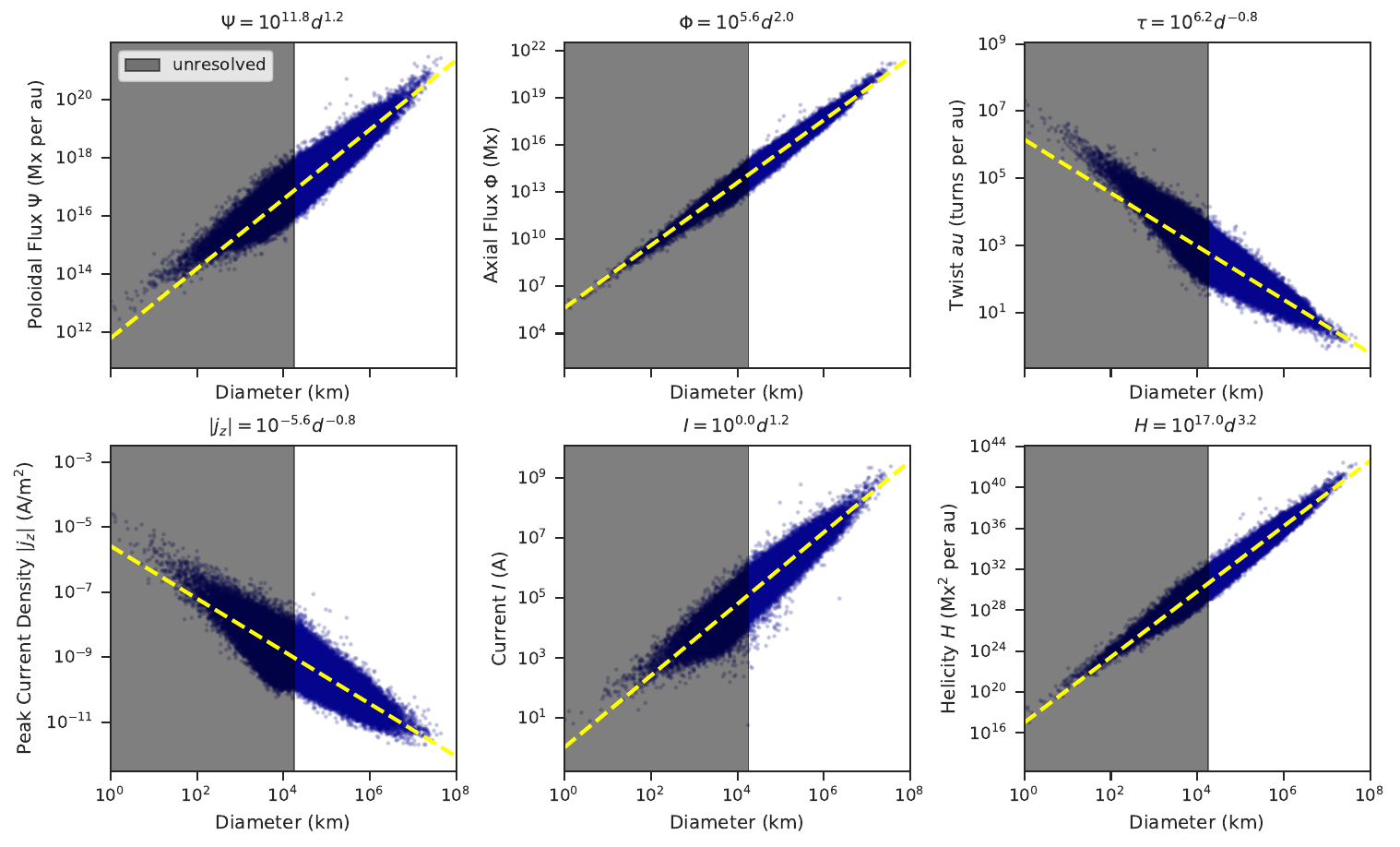}
    \caption{Scatter plots of $\Psi$, $\Phi$, $\tau$ as functions of diameter.
                The shaded region, with cutoff at $\SI{600}{\kilo\meter\per\second} \times \SI{30}{\second}$, indicates where the smallest sliding windows cannot detect SMFRs of that size for orientations too far from radial, so the reconstructions and derived parameters are less reliable in this range.}
    \label{fig:flux}
\end{figure}

Using the reconstructed cross sections from the output of our improved detection algorithm, we have access to more quantitative information on the detected SMFRs than previous studies.
Because magnetic flux is conserved under ideal MHD,
it is useful to view the axial and poloidal flux in particular.
We estimated the axial flux $\Phi$ by integrating $\int B_z dxdy$ over the closed region of the recovered cross section
(Appendix~\ref{sec:core_region_algo})
and the poloidal flux per unit length $\Psi$ as the difference between the maximum and minimum values of $A$ in the closed region.
Using these two, we estimated the average twist (number of turns turns per unit length) as $\tau = \Psi/\Phi$
(which is the equation for the twist of a cylindrical flux rope having uniform twist).
Additionally, we calculated the 2.5D helicity (per unit length) as $H = \int (A - A_{0}) B_z dx dy$
where $A_{0}$ is the the value of $A$ at the outermost field closed transverse field line and the integration is over
the closed region of the flux rope
\citep{hu_evolution_1997}.

Figure~\ref{fig:flux} shows how these parameters vary as a function of flux rope diameter.
We only consider the range of diameters that are fully resolved in terms of axial orientation
and thus have more reliable reconstructions.
From here, it appears that the axial flux is directly proportional to the area, suggesting that the axial field strength is independent of the flux rope scale.
However, although the poloidal flux is related to the diameter, it is not directly proportional. Instead, it is related with a power law index of 1.2.
These relationships mean that the axial magnetic field strength is largely independent of the flux rope size,
whereas larger flux ropes have stronger poloidal magnetic field strength on average.
The power law indices appear to remain the same regardless of solar activity level or the yearly average IMF strength (not shown), although both the poloidal and axial flux distributions appear to
shift proportionately to changes in average IMF strength.

We also calculated the peak axial current density $j_z$ from the polynomial fit to $P_t'(A)$.
Figure~\ref{fig:flux} shows that the peak axial current density shares the same power law as the twist and that the total current shares the same power law as the poloidal flux.
The current density decreases with size, whereas the total current increases with size.
\begin{deluxetable}{lrrrr}
\label{tab:difference}
\tablecaption{Median value of properties in different solar wind plasma types calculated using only events in the well-resolved range of diameters.}
\tablehead{\colhead{Quantity} & \colhead{Coronal Hole} & \colhead{Streamer Belt} & \colhead{Sector Reversal} & \colhead{Ejecta}}
\startdata
$\Phi/A$ & 5.07 nT & 4.84 nT & 3.88 nT & 9.66 nT \\
$2\Psi/d$ & 0.78 nT & 0.70 nT & 0.58 nT & 0.94 nT \\
$|R_w|$ & 0.60 & 0.46 & 0.29 & 0.47 \\
$\|\mathbf{v}_\mathrm{FR}\|$ & 608 km/s & 434 km/s & 348 km/s & 446 km/s\\
$\|\langle \mathbf{v}_\mathrm{sw} \rangle \|$ & 590 km/s & 426 km/s & 346 km/s & 429 km/s\\
$\langle n_p \rangle$ & 2.8 per cc & 4.6 per cc & 7.40 per cc & 5.2 per cc \\
$\langle T_p \rangle$ & 20.0 eV & 9.0 eV & 3.8 eV & 8.9 eV \\
\enddata      
\end{deluxetable}

\subsection{Plasma Type Dependence}

Table~\ref{tab:difference} shows how certain SMFR properties depend on the plasma type.
The axial and poloidal fluxes for a given size
tend to be higher in faster plasma types such as coronal hole origin plasma
and ejecta, but lower in slower plasma types such as streamer belt origin plasma
and sector reversal regions.
Conversely, this implies that larger flux ropes are more common in slower solar wind streams, consistent with previous results \citep{borovsky_flux_2008,hu_automated_2018}.
The {\alf}icity of SMFRs is significantly higher
in coronal hole origin plasma.
Flux rope speed and plasma speed are both higher in
faster solar wind types, but the average plasma speed is slightly lower in all types (even ones with low {\alf}icity)
due to the sunward velocity fluctuations in the flux rope frame.
This is consistent with the general understanding that the fast solar wind
is more {\alf}ic whereas the slow solar wind is less {\alf}ic.
Since the distribution is only slightly different for each non-ejecta
solar wind plasma type, it is natural to assume
that it is the same phenomenon that is observed in all three.
It is unclear whether the difference with SMFRs in ejecta plasma
is due to them being a different structure altogether or due to
a difference in the plasma conditions in ejecta plasma.
For example, if they all form locally through turbulence,
the plasma conditions should have a significant effect
on the properties of the SMFRs.

Another observation is that in Table~\ref{tab:difference},
the SMFRs are less dense and hotter in faster solar wind types
while being more dense and cooler in slower wind types.
If one estimates a characteristic pressure by multiplying average temperature by average density, the pressure coronal hole origin plasma is higher than the streamer belt plasma by a factor of $1.36$ and which, in turn, has a pressure higher than that of sector reversal plasma by a factor of $1.48$. These properties are consistent with the well-known properties of the
fast and slow solar wind. 
Since the faster solar wind types have higher pressure,
compression is a possible reason for the slight decrease in size for a given magnetic flux in faster types compared to slower types.

\section{Discussion and Conclusions} \label{sec:discussion}

In summary, we have developed an improved
version of the GS-based automated detection
algorithm,
demonstrated its improved speed and reliability qualitatively and quantitatively with simulated data,
and applied it to 27 years of \emph{Wind} data.
The improved performance enabled extending the
detection to a larger range of sizes
with significantly less computational resources.
The improved reliability justified
the removal of the $B > \SI{5}{\nano\tesla}$ threshold,
motivated by previous findings
that properties of SMFRs such as $B$
correspond to the surrounding solar wind.
This revealed
the statistical bias underlying previous conclusions regarding solar activity
and distance to HCS dependence. We summarize the findings as follows.
\begin{enumerate}
    \item SMFR diameter is log-normally distributed. The parameters of the log-normal distribution depend on solar wind type. Among non-ejecta solar wind types,
    the distribution extends to larger values in slower types,
    consistent with previous findings \citep{borovsky_flux_2008,hu_automated_2018}. We have shown that this may be a consequence of compression (Table~\ref{tab:difference}), although it could alternatively be related to proximity to the HCS.
    In ejecta plasma, the distribution is also log-normal but despite the relatively fast speed, its size range extends further than other types.
    \item The apparent solar activity dependence of the SMFR count was eliminated by removing the $B$ threshold or by using the yearly average as a flexible threshold instead. The remaining variation of SMFR count per year is consistent with proportionality to the average yearly velocity. This proportionality was also verified to hold at a timescale of a synodic solar rotation. The slope of the proportionality yielded an estimated radial density of ${\sim} 1$ SMFR per $10^6$ km. 
    \item The filling factor, or percentage of measurements within SMFRs, is independent of solar activity without a fixed $B$ threshold (having a constant value of approximately 35\%). With the fixed 5 nT threshold, there is a strong correlation between the yearly average solar wind $B$ and the filling factor of the SMFRs with $B$ above the threshold. With the yearly average $B$ as a flexible threshold, the filling factor is again independent of solar activity.
    \item The majority of the filling factor
    is contributed by SMFRs of diameters between $10^5$ and $10^6$ km. These do not exhibit a solar activity dependence.
    The contribution to the filling factor by the largest SMFRs, above approximately 0.01 au in diameter, do exhibit a strong
    solar activity dependence,
    suggesting that they are transient
    events.
    \item The filling factor and radial density of SMFRs are independent of solar wind type and distance to the HCS.
    This is evidenced by
    the consistency of the linear relationship
    between velocity and SMFR count
    despite the well-known fact that the HCS is nearly vertical during solar maximum, affecting the proximity to the HCS throughout a solar rotation.
    Moreover, the rate at which each solar wind type observed changes changes throughout the solar cycle,
    so a major solar activity dependence
    should have been observed if SMFRs were more common
    in a certain solar wind type.
    It is also supported by the virtual independence of the filling factor
    on the categorization of the
    solar wind type by the \citet{xu_new_2015} method
    (Figure~\ref{fig:occurrence_by_origin}).
    \item The previous finding of HCS distance dependence reported by \citet{cartwright_heliospheric_2010} and \citet{hu_automated_2018}
    was demonstrated to be a consequence of statistical bias. The reason for the peak of the distribution of distance to the nearest HCS crossing in both studies
    is that the same peak is found in the distribution of the measurements' distance to the nearest HCS crossing.
    We found that the reason that \citet{hu_automated_2018}
    found the peak was 1 day after the HCS crossing is because of the known fact that CIRs catch up to the HCS, resulting in faster solar wind about a day after HCS crossings with $B > \SI{5}{nT}$, resulting in both a higher event count (for both the new and old catalogs) and a higher filling factor (for the old catalog only, due to the 5 nT threshold).
    \item The orientation of SMFRs at all scales where the orientation can be resolved, including the largest scales, is consistent with Parker spiral alignment (Figure~\ref{fig:orientation_distr}).
    This is consistent with previous studies \citep{borovsky_flux_2008,hu_automated_2018}
    but is demonstrated in this study across a wider range
    of scales.
    \item Most SMFRs at 1 au have significant {\alf}icity (field-aligned flows) as determined by the {\wal} slope. Although \citet{gosling_torsional_2010} found that early SMFR lists did not contain many {\alf}ic events, {\alf}ic events were generally excluded by previous studies. The reason that previous applications of GS detection did not report a significant number of {\alf}ic SMFRs was that the use of the average velocity as the reference frame led to a {\wal} slope biased to 0. Applications to PSP \citep{chen_small-scale_2020,chen_small-scale_2021,chen_small-scale_2022} used the HT frame,
    which led to the correct {\wal} slope.
    Corrected calculation applied to the original catalog's events reveals that most of the events in the original catalog
    have high {\alf}icity,
    which we also validated using a reference frame-independent method \citep{chao_walen_2014}.
    \item Despite the high {\alf}icity,
    {\alf}ic SMFRs exhibit signatures inconsistent 
    with pure {\alf} waves.
    These include the nonzero change in $B$ (incompatible with small-amplitude {\alf} waves) and the anticorrelation between magnetic field strength and density (whereas {\alf} waves should have a positive correlation \citep{vellante_analysis_1987}). Additionally, the statistical properties of {\alf}ic SMFRs are consistent with those of non-{\alf}ic SMFRs,
    suggesting that they are the same phenomenon.
    \item Comparison of the {\alf}icity
    distribution at 1 au (our event list) and PSP \citep{chen_small-scale_2022}
    results in little difference between the two distributions (Figure~\ref{fig:PSPcomparison}).
    The peak of the distribution is the same.
    The PSP results have slightly less low-{\alf}icity events and slightly more high-{\alf}icity events.
    It is unclear if these disagreements are due to differences in methodology,
    radial evolution,
    or PSP spending more time in {\alf}ic soolar wind.
    \item Using the additional information provided by the new detection method,
    we found that poloidal flux $\Psi$, axial flux $\Phi$, twist $\tau$, current density $j_z$, and helicity $H$ follow power laws with respect to diameter. The axial flux power law of $\Phi \propto d^{2.0}$ implies that the average axial field strength $\avg{B_z} \propto \Phi/d^2$ is independent of size. The poloidal flux power law $\Psi \propto d^{1.2}$ implies that the average poloidal field strength $\avg{B_\phi} \propto \Psi/(2d)$ is \emph{not} independent of size, but increases slightly with size as $\avg{B_\phi} \propto d^{0.2}$. As a consequence, larger SMFRs have slightly higher total field strength.
    The helicity power law $H \propto d^{3.2}$ suggests that if the larger SMFRs form by merging of smaller SMFRs, either helicity or total area is not conserved (if both were conserved, we would have $H \propto d^{2.0}$).
    Presumably, area is less likely to be conserved. These power laws can be compared to large-scale MHD simulations in future studies.
    \item By comparing the average properties of SMFRs in different solar wind types, we confirmed that the SMFR properties closely follow the properties of the surrounding solar wind. For example, magnetic field strength and temperature are highest in coronal hole-origin plasma, whereas density is highest in sector reversal region plasma. Similarly, {\alf}icity is higher in coronal hole-origin plasma.
\end{enumerate}

SMFRs were originally seen as transient structures, but results from this and other recent studies (Section~\ref{sec:intro}) suggest that the solar wind is a sea of SMFRs.
The primary candidates for the origin of SMFRs according to the
earliest studies were relatively small CMEs \citep{feng_interplanetary_2008} or reconnection across the HCS \citep{moldwin_ulysses_1995,moldwin_small-scale_2000,cartwright_heliospheric_2010}.
However, for most SMFRs, the lack of solar activity dependence of the radial density makes it unlikely that
they are related to small CMEs.  
The complete independence of distance from the HCS suggests that it is unlikely that
reconnection across the HCS is a major source of SMFRs, either.
Of course, it is likely that both of these mechanisms contribute sub-populations,
since some solar eruptions have been directly linked to SMFRs at 1 au \citep{rouillard_solar_2011}.
These SMFRs would be transients as opposed to filling the solar wind.
However, if one is to study them, it is not sufficient to simply look for long events or events with elevated $B$,
as most of those may just be particularly large fluctuations, the tails of the log-normal distributions of
the main population SMFRs' size and magnetic field strength.
Additional factors such as having significantly different physical properties from surrounding solar wind should be considered.
In fact, since SMFRs above ${\sim}0.01$ au do appear to have a
strong solar cycle dependence (Figure~\ref{fig:yearly_variation} (c)),
this may be possible to use as a threshold to identify transient SMFRs.
However, unlike ICMEs, these large SMFRs are aligned with the Parker spiral, just like smaller SMFRs (Figure~\ref{fig:orientation_distr}).

Within the sea of flux ropes model, the origin and dynamics of the flux ropes remain contested. \citet{borovsky_flux_2008} suggests that solar wind flux tubes typically do not interact through reconnection due to the expansion of the solar wind. However, others believe that reconnection processes result in constant destruction, creation, and merging. \citet{greco_waiting-time_2009} and many others have demonstrated how MHD turbulence can generate flux ropes with waiting times similar to those observed in situ between current sheets. This is often cited as evidence of local generation of the current sheets via turbulence, but it can result from repeated reconnections causing random multiplications of structure sizes \citep{matthaeus_low-frequency_1986}, and this process can alternatively happen at the Sun.
The non-Gaussian current density distribution observed by \citet{greco_waiting-time_2009} was found to be similar to the in-situ SMFRs by \citet{zheng_observational_2018}, which may be evidence of local generation via turbulence, but it is in fact not independent evidence from the log-normal distribution of size, since current density is related to size by a power law. 
As of now, it is unclear whether the structures are generated locally or originate as structures from the Sun (in which case, it is unclear to what extent they evolve through reconnection processes between the Sun and 1 au).

We have shown that using the correct calculation of the {\wal} slope,
the {\alf}icity of SMFRs tends to be high even at 1 au.
The similarity of the {\wal} slope distribution between the PSP results and the 1 au results (Figure~\ref{fig:PSPcomparison})
is interesting and requires further research.
However, they usually contain what appear to be embedded {\alf} waves.
Relative to its average plasma velocity, when field aligned flows are present, the flux rope structure tends to propagate outward (antisunward)
along the Parker spiral at slightly less than the {\alf} speed; equivalently, in the flux rope frame of reference,
there is a sub-{\alf}ic sunward plasma flow.
The result of antisunward propagation was
found based on HT analysis in previous studies of SMFRs observed by PSP \citep{chen_small-scale_2022}
and of general {\alf}ic magnetic structure in the solar wind \citep{borovsky_motion_2020}.
\citet{paschmann_discontinuities_2013} also
found that directional discontinuities, both rotational and tangential,
propagate antisunward based on electron strahl measurements combined with HT analysis.
Considering the abundance of SMFRs, it is likely that most directional discontinuities
are related to SMFRs.
Similarly, while fluctuations in the solar wind are commonly attributed to outward-propagating {\alf} waves
\citep{belcher_large-amplitude_1971}, SMFRs appear to be uniformly present in all solar wind types, {\alf}ic or not. 
Thus, a significant portion of, if not the majority of, the {\alf} waves at 1 au appear to be embedded within SMFRs.

What causes the embedded {\alf}ic flows in SMFRs?
\citet{borovsky_plasma_2020} mentioned that perturbations perpendicular to a flux tube's axis propagate along the axis
relative to the plasma at a speed related to the {\alf} speed (see references therein).
Such a propagating disturbance is essentially a torsional {\alf} wave.
Within the model of fossil structures connected to the Sun,
they suggested that the perturbations could be due to the
shuffling of flux tubes at the Sun.
This is a plausible explanation for the common field-aligned flows that are more or less the same
at 1 au and PSP.
However, field-aligned flows can also occur through local processes
(see, for example, the field-aligned flows in the benchmark MHD simulation in Figure~\ref{fig:sim_examples}).
Whether the perturbations originate from the Sun or throughout the solar wind, one would still expect them to mostly propagate away from the Sun because of the super-{\alf}ic speed of the solar wind.

The structure of SMFRs requires the magnetic field to rotate about a central axis. These deflections
could be associated with switchbacks (SBs), the study of which
has become quite popular due to their prominence in PSP observations
\citep{bale_highly_2019,kasper_alfvenic_2019}.
As for SMFRs, numerous mechanisms have been proposed
to explain the generation of SBs
(e.g., \citet{squire_-situ_2020,ruffolo_shear-driven_2020,drake_switchbacks_2021,huang_statistical_2023}).
Some mechanisms propose a solar origin, while others propose a local origin.
Due to the numerous plausible origin mechanisms,
it is likely that a number of mechanisms
have varied levels of contributions at different distances
from the Sun.
Observationally, \citet{pecora_magnetic_2022}
showed that the occurrence of SBs
per unit length decreases sharply within 0.2 au of the Sun,
whereas it increases gradually beyond 0.2 au,
implying that local dynamics play an important role.
\citet{drake_switchbacks_2021} discussed
how flux ropes can appear as switchbacks in PSP
observations.
By including SMFRs with high {\alf}icity,
\citet{chen_small-scale_2021,chen_small-scale_2022}
demonstrated that many SBs observed by PSP are related to SMFRs.
We have found that SMFRs at 1 au have essentially the same {\alf}icity as SMFRs observed by PSP,
which means that they can also appear as SBs even if {\alf}icity is required.

Interestingly, as we have found to be the case with SMFRs, SB occurrence is correlated with bulk velocity \citep{mozer_origin_2021,jagarlamudi_occurrence_2023}.
We have shown that the occurrence of SMFRs is correlated with the average bulk velocity as a consequence of a constant radial density. If we find that SBs, too, have constant radial density,
that would be additional observational evidence that they are rotations of the magnetic field
caused by structures such as SMFRs.
However, evaluating precisely whether the radial density of SBs is constant will be challenging
because the types of solar wind and levels of solar activity observed by PSP are limited, and there can be noticable fluctuations in SMFR counts even on the timescale of a solar rotation (Figure~\ref{fig:yearly_variation}). Nevertheless, this is an interesting work to be carried out in the future.

Future studies of SMFRs should take into consideration some of the points we have raised regarding
the detection of SMFRs and the statistical interpretation of the results:
\begin{enumerate}
\item Variation in the number of events must be interpreted with caution.
A fixed threshold on $\avg{B}$ can cause significant statistical bias because not all SMFRs are transients.
In MHD simulations and observations, the SMFR $B$ rarely differs significantly from the surrounding plasma.
Removing the threshold only increases the number of events by approximately a factor of two,
but it completely changes the conclusions regarding solar activity dependence.
The distribution of distance to large-scale structures such as the HCS
needs to be compared with the overall measurements' distribution.
It may be that events appear to be close to or far from a given large-scale structure,
but in fact they share the same distribution of distance to the structure as arbitrarily chosen points in time.
\item {\alf}icity, or correlation between changes in velocity and magnetic field, can be measured through the slope of the {\wal} relation $\delta \vb{v} \equiv \vb{v} - \vHT \propto \vb{v}_A$. It must be evaluated in the HT frame \citep{khrabrov_dehoffmann-teller_1998},
which differs from $\avg{v}$ when $\avg{\delta \vb{v}} \neq 0$.
Averaging over a single flux rope
will result in a scatter plot where all of the components are centered on the origin, resulting
almost invariably in a {\wal} slope of zero.
Averaging over a longer interval may be less problematic,
but because the {\alf}ic disturbances have a preferential direction (antisunward),
it may still give inaccurate results.
Alternatively, the {\wal} slope may be evaluated using a
frame-independent method \citep{chao_walen_2014}, which produces the same statistical results as the HT frame (Figure~\ref{fig:original_catalog_walen_distr}).
\item A strong field-aligned {\alf}ic flow does not necessarily mean an event candidate is a pure {\alf} wave.
From a theoretical point of view, {\alf}ic flows are expected to be observed in flux ropes, since there are many processes that can cause them (see \citet{gosling_torsional_2010} and references therein).
Other factors must be considered to distinguish pure {\alf} waves from SMFRs,
such as a nonzero change in magnetic field strength or anticorrelation between magnetic field strength and density.
\end{enumerate}

A major but necessary limitation of this study is that the events
are detected based on a single spacecraft. A single spacecraft cannot
measure the gradient of the magnetic field,
so the 2D assumption that the GS method is based on cannot be directly validated. Even if the 2D assumption is correct
and there are flux ropes present,
if the wrong boundaries are selected,
the reconstructed cross section will be totally inaccurate,
as demonstrated in Section~\ref{sec:benchmark}.
Nevertheless, the usage of single-spacecraft data
was necessary for the long period of time that it afforded,
and our benchmarking against an MHD simulation suggests that
most of the detected events are flux ropes even if the reconstruction is inaccurate,
especially those with more than 30 data points (Table~\ref{tab:comparison}).
While only single spacecraft measurements are available to use
for such a large-scale statistical study as this one,
missions such as MMS can be used to determine the reliability 
from single spacecraft detection.
In a forthcoming study, we will use a novel
GS-inspired technique to detect and reconstruct
SMFRs from MMS data
to validate the findings in this and other single spacecraft studies.

The source code for the new detection algorithm is
available at \url{https://github.com/hafarooki/PyMFR}.
For an easy-to-read CSV file containing basic information
about each event,
as well as plots for each event, see \url{https://doi.org/10.6084/m9.figshare.24547810}.
For the version of the code used in this paper, together with the code for downloading and processing data,
running the detection algorithm, and generating figures, along with the data used, see
\url{https://doi.org/10.6084/m9.figshare.24547798}.
There is much more insight to be gained from the new database in future studies.
For example, why is the {\wal} slope similar for flux ropes near the Sun and at 1 au, having the same peak in the distribution? Can the power law for poloidal and axial magnetic flux
help to determine the origin of the SMFRs? Are all of the observed statistical properties of SMFRs compatible with
results from MHD simulations? What other information can be derived from this study's novel database
containing on the order of $10^5$ flux rope cross sections derived from GS reconstruction? These questions should be investigated by forthcoming studies.

\section{Acknowledgements}

We thank the teams at NASA for providing the \emph{Wind} spacecraft data used for this study.
We gratefully acknowledge support from NASA grant 80NSSC20K1282 and NSF grants AGS-2229064/2229065,
OAC-2320147, and OPP-2032421.
J.L was supported by NSF grant, AGS 2114201.
F. Pecora is supported by PSP HelioGI under grant number 80NSSC21K1765 at the University of Delaware.
The simulations have been performed at the Newton cluster at University of Calabria and the work is supported by ``Progetto STAR 2-PIR01 00008'' (Italian Ministry of University and Research). S. Servidio acknowledges supercomputing resources and support from ICSC–Centro Nazionale di Ricerca in High Performance Computing, Big Data and Quantum Computing–and hosting entity, funded by European Union–NextGenerationEU.
H. Farooki thanks 
Qiang Hu,
Lynn B. Wilson, 
Nada al-Haddad,
Robert T. Wicks,
Ben Lynch, 
Bill Matthaeus,
Paul Cassak, 
Jimmy Juno, 
Charles Farrugia,
Kyung-Eun Choi,
Kristopher Klein,
Joe Borovsky,   
Brian Welsch,
Laxman Adhikari, 
Sanchita Pal, 
No$\acute{\text{e}}$ Lugaz,
Wenyuan Yu, 
and Phil Isenberg
for helpful comments and
discussions.

\appendix
\section{Theoretical Background} \label{sec:theory}

\subsection{The Grad-Shafranov (GS) Technique}

For any 2.5D magnetic field ($\pdv*{}{z} = 0$ but $B_z(x, y) \neq 0$),
we can write the magnetic field in terms
of a magnetic vector potential $\vb{A}$ as $\vb{B} = \curl{\vb{A}} = \pdv{A}{y} \vu{x} - \pdv{A}{x} \vu{y} + B_z(x, y) \vu{z}$
where $A \equiv \vb{A} \vdot \vu{z}$.
The significance of $A$ can be seen by taking its gradient
along the magnetic field direction:
$\grad{A} \vdot \vb{B} = \pdv{A}{x} \pdv{A}{y} - \pdv{A}{y} \pdv{A}{x} = 0$.
This implies that $A$ is a field line invariant
(constant along transverse field lines). Thus the contours of $A$ in the plane perpendicular to $\vu{z}$
(isosurfaces of $A$ when viewed in 3D)
are transverse (in-plane) magnetic field lines.
Using Ampere's law, $\mu_0 j_z = \vu{z} \vdot \curl{\vb{B}} = \pdv*{B_y}{x} - \pdv*{B_x}{y} = -\pdv*[2]{A}{x} - \pdv*[2]{A}{y} = - \laplacian{A}$,
so the source of $A$ is the axial current density:
\begin{equation} \label{eq:Asource}
    \laplacian{A} = -\mu_0 j_z
\end{equation}

Evaluating the flux rope velocity $\vFR$
requires finding the frame of reference
in which the magnetic structure does not change, i.e. $\pdv*{\vb{B}}{t} = -\curl{\vb{E'}} = 0$, where $\vb{E'} = \vb{E} + \vFR \cross \vb{B}$.
A sufficient (but not necessary) condition for this is simply that $\vb{E} = -\vFR \cross \vb{B}$ and $\vb{E'} = 0$.
Such a reference frame is called a deHoffman-Teller or HT frame, denoted $\vHT$ \citep{de_hoffmann_magneto-hydrodynamic_1950}.
An optimal HT frame can be found efficiently using a linear algebra technique \citep{khrabrov_dehoffmann-teller_1998}.
Typically, $\vHT$ is found assuming that the solar wind electric field is $\vb{E} = -\vb{v} \cross \vb{B}$,
since we operate at MHD scales and direct measurements
of the electric field are not always available.
Furthermore, the proton velocity is usually used to calculate $\vb{E}$,
although if quality electron velocity measurements are available, they could provide more reliable results \citep{khrabrov_dehoffmann-teller_1998,puhl-quinn_systematics_2000}.
A valid HT frame is usually present in the solar wind,
so $\vFR \equiv \vHT$.
The power of the HT frame is that it provides a strong validation of the structure being time-static, since $\vb{E'} = 0$ is a sufficient condition for $\pdv*{\vb{B}}{t} = 0$.

If we find a valid $\vHT$, we know that the structure is approximately static over time.
If $\vb{v}(t) = \vHT$, we can assume magnetostatic equilibrium ($\vb{j} \cross \vb{B} = \grad{p}$).
It follows that $j_x B_y - j_y B_x = \pdv*{p}{z} = 0$,
so $j_x/j_y = B_x/B_y$. This is equivalent to saying that $B_z$ is a field
line invariant for magnetostatic structures, because
we also have that $\mu_0 j_x = \pdv*{B_z}{y}$
and $\mu_0 j_y = -\pdv*{B_z}{x}$,
so $\vb{B} \vdot \grad{B_z} = -\mu_0 (B_x j_y - B_y j_x)$.
Moreover, $p$ is a field line invariant, since
$\vb{B} \vdot \grad p = \vb{B} \vdot (\vb{j} \cross \vb{B}) = 0$.
Assuming each transverse field line has a unique value of $A$, $B_z = B_z(A)$ and $p = p(A)$. It can be shown that $j_z = \dv*{}{A} \qty[p + B_z^2/2\mu_0]$.
The quantity $p + B_z^2/2\mu_0$ is commonly called the transverse pressure, denoted $P_t$.
If $\vu{z}$ is known, $P_t(A)$ can be fitted from spacecraft
measurements with an appropriate function, and the derivative
can be used as $j_z$ and the source term for $A$.
The structure in the magnetically connected region above and below the spacecraft path can be recovered by solving
the GS equation as an initial value problem,
a process known as GS reconstruction \citep{sonnerup_grad-shafranov_2006}.

The original GS equation and reconstruction assume a magnetostatic structure
with no flow in the $\vFR$ frame. However, the HT frame is still present, and a plasma structure can still be in a stationary state if there are finite, field-aligned flows. \citet{sonnerup_grad-shafranov_2006} derived a GS-like equation and a reconstruction process for such a scenario, where the remaining flow can be written in terms of the {\alf} speed as $\Delta \vb{v} \equiv \vb{v} - \vFR = M_A(x, y)\vb{v}_A$,
where $M_A \equiv \Delta v/v_A$ is the {\alf} Mach number in the frame $\vFR$
(not the spacecraft frame).
In the general case, $B_z$ and $M_A$ are not field line invariants,
but $(1 - M_A^2)B_z$ is.
For the special case where $M_A \equiv M_A(A)$ is a field line invariant, \citet{teh_gradshafranov_2018} introduced a further simplified equation
in terms of $A' \equiv (1 - M_A^2)A$ rather than $A$.
It is especially simplified if $M_A = \mathrm{constant}$, in which case $(1 - M_A^2)$ can be factored out, and the equation becomes:
\begin{equation} \label{eq:modifiedGS}
\laplacian{A} = -\mu_0 \frac{d}{dA} \qty[\frac{B_z^2}{2\mu_0} + \frac{1}{1 - M_A^2} p + \frac{M_A^2}{1 - M_A^2} \frac{B^2}{2\mu_0}] = -\mu_0 j_z \impliedby \mathrm{Eq~\ref{eq:Asource}}
\end{equation}
so that the approach is essentially the same as the original GS method except $j_z = \dv*{(B_z^2/2\mu_0 + p)}{A}$ must be replaced with $j_z = \dv*{(B_z^2/2\mu_0 + p/(1 - M_A^2) + \frac{M_A^2}{1 - M_A^2} B^2/2\mu_0)}{A}$. This special case
has become very important in SMFR studies
because recent studies found that closer to the Sun, there are few static SMFRs
but still many with field-aligned flows.
Observationally, SMFRs with field-aligned flows
appear to have a constant $M_A$ that can be estimated as the {\wal} slope $R_w$ (the slope of a linear fit through the origin $\vb{v} - \vHT = R_w \vb{v}_A$; it is sometimes estimated using a general linear fit, but we use the linear fit through the origin)
\citep{chen_small-scale_2021,chen_small-scale_2022}. As we showed,
SMFRs with field-aligned flows are dominant at 1 au, not just near the Sun.

\subsection{Application to Spacecraft Measurements of Flux Ropes}

\begin{figure}
    \centering
    \resizebox{6cm}{!}{%
    \begin{tikzpicture}
        \coordinate (O) at (0,0,0);
        \coordinate (A) at (-1,0,0);
        \coordinate (B) at (0,1,0);
        \coordinate (C) at (0,0,-1);
        \coordinate (D) at (1.5,0,3.5);
        \node[cylinder, draw, shape aspect=3, rotate=-135, minimum height=4cm, minimum width=.75cm, cylinder uses custom fill, cylinder end fill=black!00, cylinder body fill=black!20, opacity=0.5]{};
        \draw[-latex, red] (O) -- (A) node[below] {$\vu{x}$};    
        \draw[-latex, green] (O) -- (B) node[above] {$\vu{y}$};
        \draw[-latex, blue] (O) -- (C) node[below] {$\vu{z}$};
        \draw[-latex, magenta] (O) -- (D) node[below] {$\vFR$};
    \end{tikzpicture}
    }
    \caption{Flux rope coordinate system. $\vu{z}$ is the cylindrical direction along which $\pdv*{}{z} = 0$. The perpendicular plane is divided into $\vu{x}$ and $\vu{y}$, defined so that the spacecraft only moves through the cross section along $\vu{x}$ minus the motion along $\vu{z}$. Therefore, the velocity of the flux rope in the spacecraft frame ($\vFR$; assumed constant) is contained in the xz plane. Since the spacecraft moves along $\vu{x}$ within the cross section, the x component of $\vb{v}_{FR}$ must be negative, since $\vb{v}_{FR}$ is measured in the spacecraft frame of reference.}
    \label{fig:coords}
\end{figure}
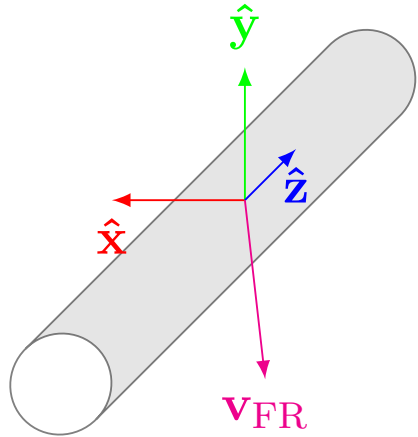

The GS technique (summarized above) requires a coordinate system where $\pdv*{}{z} = 0$.
It is common to define a coordinate system (Figure~\ref{fig:coords})
such that $\vu{z}$ is the cylindrical axis,
the spacecraft moves in the $\vu{x}$ direction
through the cross section (so that $\vb{v}_{FR} \vdot \vu{x} < 0$ since $\vFR$ is in the spacecraft's frame of reference), and $\vu{y}$ is the perpendicular
direction in the cross section defined so $x = x(t)$
but $y = y_0$ (the spacecraft does not move in the y direction).
To construct the coordinate system, it is sufficient to find
$\vu{z}$ and $\vFR$. From there, $\vu{x} = -\mathrm{normalize}(\vFR - \vu{z}(\vFR \vdot \vu{z}))$
and the right-hand rule specifies $\vu{y} = \vu{z} \cross \vu{x}$. 

Finding $\vu{z}$ is essential but challenging.
Using measurements from only a single spacecraft,
the gradients of the magnetic field components are not specified, so $\pdv*{}{z}$ cannot be directly verified by any sufficient condition.
For flux ropes observed by a single spacecraft, \citet{hu_reconstruction_2002} introduced a method to determine $\vb{z}$.
In a (reasonably simple) flux rope structure, each transverse field line has a unique value of $A$ (and other field line invariants, such as $P_t$). Each field line (thus each value of $A$)
is observed twice as a spacecraft passes through the flux rope,
so $P_t = P_t(A)$.
If a given $\vb{z}$ is the correct orientation,
then the derived $A$ versus the derived $P_t$ should show minimal scatter since $P_t$ should be a single-valued function of $A$. Thus $\vb{z}$ is selected to minimize scatter between $P_t$ and $A$, quantified as the difference residue $R_\mathrm{diff}$ described in \citet{hu_reconstruction_2002}. $R_\mathrm{diff}$ is essentially the root mean square difference between each value of $P_t$ and the corresponding value interpolated to match the same value of $A$ from the other side of the measurement interval, normalized by the range of $P_t$ to avoid selecting an orientation where $P_t = \mathrm{constant}$, which would imply zero current density.
(Note that \citet{hu_reconstruction_2002} actually used evenly spaced $A$ values to get interpolated $P_t$ values from either side for calculating $R_\mathrm{diff}$ whereas \citet{hu_automated_2018} compared each measured value to the corresponding value from the other side. The advantage of the latter approach is that it gives less bias to measurements that happen to have a large spacing in $A$, which is common at the flux rope boundary. We use the latter approach. Also, \citet{hu_automated_2018} added a factor of $1/\sqrt{2}$, but we did not use it.)
This approach usually leads to a well-determined flux rope orientation, although there is much uncertainty in distinguishing $\vu{x}$ from $\vu{z}$, especially if $B_x$ is symmetric \citep{hu_reconstruction_2002}.
Besides $R_\mathrm{diff}$, the validity of the polynomial fit to $P_t(A)$
required for GS reconstruction is validated using a similar quantity $R_\mathrm{fit}$ (introduced by \citet{hu_multiple_2004}) which is equivalent to the root mean squared difference between the measured $P_t(t)$ and the fitted $P_t(A)$
normalized by the range ($\max(P_t) - \min(P_t)$).

In principle, evaluating $A(t)$ would require integrating
$dA = (\pdv*{A}{t})dt + (\pdv*{A}{x}) dx + (\pdv*{A}{y}) dy$.
However, the analysis is limited to time-static structures, so $\pdv*{A}{t} = 0$.
As the coordinate system is defined so that $\vFR$ is contained in the x-z plane, $dx = -v_{x,\mathrm{FR}} dt$ (it is assumed that the spacecraft is sitting still as the flux rope passes through) and $dy = 0$.
Therefore, once the coordinate system in terms of the measurement coordinate system is known, $A$ along the spacecraft path is given by:
\begin{equation} \label{eq:Astrip}
    A(x, y_0) = \int_{x_0}^x - B_y dx = -|v_{x,\mathrm{FR}}| \int_{t_0}^t B_y dt
\end{equation}

\subsection{Automated Detection of Flux Ropes from Spacecraft Measurements}

The main signature of a flux rope is the bipolar $B_y$ which crosses the origin once (corresponding to the field lines pointing up on one side of the cross section and down on the other)
and $B_z$ increasing towards the center (under force-free conditions, this is necessary considering its relation to the current density).
However, this signature is not visible in every coordinate system. It is often the case that in the spacecraft measurement coordinate system there is no bipolar component of the magnetic field.
As a result, a simple visual inspection of timeseries measurements will miss most of the flux ropes.

\citet{hu_automated_2018} developed an automated detection algorithm and applied it to
build a catalog of SMFRs from 21 years (1996-2016) of \emph{Wind} data.
Their algorithm applies a sliding window to the spacecraft measurements
and tests the hypothesis that a given interval is a flux rope.
Thus, their algorithm is an exhaustive search algorithm.
After finding the windows that are acceptable flux rope candidates, the overlapping candidates are cleaned,
and the gaps between the detected events are filled with events detected
using smaller sliding windows. They used sliding windows that ranged from about 6 hours to 10 minutes.
For each interval, the algorithm searches the entire $4\pi$ space
with coarsely separated axial orientations ($\SI{20}{\degree}$ azimuthal, $\SI{10}{\degree}$ latitudinal) for the $\vu{z}$ that minimizes $R_\mathrm{diff}$.
The algorithm determines $\vFR$ and uses the test $\vu{z}$ to set up the flux rope coordinate system.
Using the $\vu{y}$ given by the test $\vu{z}$ and $\vFR$, $A$ is calculated
using Equation~\ref{eq:Astrip}.

The original detection algorithm considers every possible interval with the $\vu{z}$ that minimizes $R_\mathrm{diff}$.
To be considered a flux rope, a given interval with its optimal $\vu{z}$ must have (1) a derived $A(x, y_0)$ with a single stationary point (a necessary condition for each transverse field line being crossed twice) that can be trimmed to the boundaries ($A_f = A_0$) without shortening the interval to the next sliding window length (with 5-minute spacing); (2) $R_\mathrm{fit} < 0.14$ and $R_\mathrm{diff}/\sqrt{2} < 0.12$ (with $R_\mathrm{diff}$ being scaled down by a factor of $1/\sqrt{2}$ to make it comparable to $R_\mathrm{fit}$); (3) no strong plasma flow in the $\vFR$ frame ($R_w < 0.3$); (4) the peak $A$ corresponds to the peak $P_t$ ($P_t$ at the peak must be in the top 15\%); and (5) a relatively high average magnetic field strength $\langle B \rangle > \SI{5}{\nano\tesla}$ to exclude small fluctuations.

Thus, in summary, the algorithm generates a list of nonoverlapping intervals
containing time-static magnetic fields satisfying the hypothesis of being
2.5D magnetostatic structures where each field line is crossed twice and having a strong axial current density and magnetic field strength. GS reconstruction of the detected events usually (but not always) reveals the presence of a flux rope, although the impact parameter is often so high that none of the field lines that are closed in the map cross the spacecraft path, hinting at significant uncertainty. It is worth emphasizing that the original algorithm does not perform GS reconstruction on the detected events; rather, the significant current density and crossing of field lines twice is assumed to correspond to a flux rope.

\section{Algorithm to Find Core Closed Region} \label{sec:core_region_algo}

To find the core region with closed transverse field lines in a reconstructed map $A(x, y)$,
we introduce the following procedure. First, we normalize $A(x, y)$ so that $A(x, y_0)$ (where the line segment $y = y_0$ is the observed strip of the interval) starts at 0, peaks at $+1$, then goes back to 0. Then, we construct another map called \texttt{visited}, which starts with all zeros except at the peak position along $y_0$, which is initialized to 1. Then, we run the following in a repeating loop:
For each pixel $x, y$, retrieve the largest neighboring value in a 3x3 square $\max(\texttt{visited}(x\pm1, y\pm1))$. Where $\max(\texttt{visited}(x\pm1, y\pm1)) > A(x, y)$, update $\texttt{visited}(x, y) \to A(x, y)$. Where $\max(\texttt{visited}(x\pm1, y\pm1)) < A(x, y)$, update $\texttt{visited}(x, y) \to \max(\texttt{visited}(x\pm1, y\pm1))$ unless $\max(\texttt{visited}(x\pm1, y\pm1)) = \max(\texttt{visited})$ and $A(x, y)$ has the greatest value out of those pixels whose neighbor is the current maximum $\texttt{visited}(x, y)$, in which case update $\texttt{visited}(x, y) \to A(x, y)$. $\texttt{visited}$ is only ever updated where the change would increase its value. 
When there are no further changes, the loop ends.
At the end, there is a single peak of $\texttt{visited}$ which is the peak connected to the observed peak $A(x, y_0)$ by following the steepest increase of $A(x, y)$ from the observed peak.
For the monotonically decreasing region around the peak, $\texttt{visited}(x, y) = A(x, y)$. If there is a region where $A(x, y)$ starts increasing again, then $\texttt{visited}(x, y) < A(x, y)$ there. Hence the largest closed transverse field line is equivalent to the contour around the region of $\texttt{visited}$ that is greater than the largest value of $\texttt{visited}$ at the boundaries of the reconstructed map,
or the largest value of $\texttt{visited}(x, y)$ where $\texttt{visited}(x, y) < A(x, y)$, or 0, whichever is greatest. We have tested this procedure and found that it successfully identifies the closed flux rope region contained in the map of $A(x, y)$.

\bibliography{main.bib}{}
\bibliographystyle{aasjournal}

%% This command is needed to show the entire author+affiliation list when
%% the collaboration and author truncation commands are used.  It has to
%% go at the end of the manuscript.
%\allauthors

%% Include this line if you are using the \added, \replaced, \deleted
%% commands to see a summary list of all changes at the end of the article.
%\listofchanges

\end{document}